\documentclass[nofootinbib,floatfix,onecolumn,superscriptaddress,a4paper,prl]{revtex4}
\pdfoutput=1
\usepackage{amsmath,amsfonts,amssymb,epsfig}
\usepackage{slashed}
\usepackage{listings}
\numberwithin{equation}{section}

\usepackage{graphicx}
\usepackage{cancel}

\usepackage{color}
\usepackage{ulem}

\usepackage{hyperref}

\usepackage{enumerate}

\usepackage{xspace}

\newcommand{\SARAH}{{\tt SARAH}\xspace}
\newcommand{\SPheno}{{\tt SPheno}\xspace}
\newcommand\Mathematica{{\tt Mathematica}\xspace}
\newcommand{\Fortran}{\texttt{Fortran}\xspace}

\newcommand{\exclude}[1]{}
\def\nn{\nonumber}

\def\beq{\begin{equation}}
\def\eeq{\end{equation}}
\def\bal{\begin{align}}
\def\eal{\end{align}}

\def\s2b{s_{2\beta}}
\def\c2b{c_{2\beta}}

\def\2b2[#1,#2][#3,#4]{\left( \begin{array}{cc} #1 & #2 \\ #3 & #4 \end{array}
\right)}
\def\3b3[#1,#2,#3][#4,#5,#6][#7,#8,#9]{\left( \begin{array}{ccc} #1 & #2 &#3 \\
#4 & #5 & #6\\#7&#8&#9\end{array} \right)}
\def\thv[#1,#2,#3]{\left( \begin{array}{c} #1 \\ #2 \\ #3 \end{array} \right)}
\def\twv[#1,#2]{\left( \begin{array}{c} #1 \\ #2 \end{array} \right)}

\def\twomat[#1,#2][#3,#4]{\left( \begin{array}{cc} #1 & #2 \\ #3 & #4 \end{array} \right)}
\def\threemat[#1,#2,#3][#4,#5,#6][#7,#8,#9]{\left( \begin{array}{ccc} #1 & #2 & #3\\ #4 & #5 & #6 \\ #7 & #8 & #9 \end{array} \right)}
\def\twovec[#1,#2]{\left( \begin{array}{c} #1  \\ #2 \end{array} \right)}

\newcommand{\TeV}{\text{TeV}}
\newcommand{\qref}[1]{Eq.~(\ref{#1})}

\def\ov{\overline}

\def\beq{\begin{equation}}
\def\eeq{\end{equation}}
\def\bea{\begin{eqnarray}}
\def\eea{\end{eqnarray}}

\newcommand{\propS}{{\rm S}}
\newcommand{\propU}{{\rm U}}
\newcommand{\propM}{{\rm M}}
\newcommand{\propV}{{\rm V}}
\newcommand{\propW}{{\rm W}}
\newcommand{\propX}{{\rm X}}
\newcommand{\propY}{{\rm Y}}
\newcommand{\propZ}{{\rm Z}}
\def\lagr{{\cal L}}

\def\cc{{\mathrm c.c.}}
\newcommand{\re}{\text{Re}}

\newcommand{\lnbar}{{\overline{\rm ln}}}
\newcommand{\Fbar}{\overline{F}}

\newcommand{\AddrLPTHE}{%
1-- Sorbonne Universit\'es, UPMC Univ Paris 06, UMR 7589, LPTHE, F-75005, Paris, France \\
2-- CNRS, UMR 7589, LPTHE, F-75005, Paris, France 
}

\newcommand{\AddrCERN}{%
Theory Division, CERN, 1211 Geneva 23, Switzerland}

\newcommand{\AddrBonn}{%
Bethe Center for Theoretical Physics \& Physikalisches Institut der Universit\"at Bonn\\Nu{\ss}allee 12, 53115 Bonn, Germany}

\begin{document}
\hfill{BONN-TH-2015-04, CERN-TH-2015-044}

\title{Two-loop Higgs mass calculation  from a diagrammatic approach}

\author{M.\ Goodsell} \email{goodsell@lpthe.jussieu.fr}\affiliation{\AddrLPTHE}
\author{K.\ Nickel} \email{nickel@th.physik.uni-bonn.de}\affiliation{\AddrBonn}
\author{F.\ Staub}\email{florian.staub@cern.ch}\affiliation{\AddrCERN}

%\keywords{}

%\pacs{??, ??, ??}

% 
\begin{abstract}
We calculate the corrections to the Higgs mass in general theories restricted to the case of massless gauge bosons (the gaugeless limit). We present analytic expressions for the two-loop tadpole diagrams, and corresponding expressions for the zero-momentum limit of the Higgs self energies, equivalent to the second derivative of the two-loop effective potential. We describe the implementation in \SARAH, which allows an efficient, accurate and rapid evaluation for generic theories. In the appendix, we provide the expressions for tadpole diagrams in the case of massive gauge bosons.
\end{abstract}

\maketitle

\section{Introduction}
With the discovery of the Higgs boson in the range of 125 -- 126~GeV the standard model (SM) has been completed \cite{Chatrchyan:2012ufa,Aad:2012tfa}. The uncertainty in the Higgs mass measurement has continuously decreased and is well below 0.5~GeV today \cite{Khachatryan:2014ira}. This small uncertainty is currently much better than the theoretical prediction in any scenario beyond the Standard Model (BSM). Therefore, more precise calculations are necessary to better confront BSM models with the Higgs mass measurement. This has two motivations of particular weight: (1) In the minimal supersymmetric standard model (MSSM) and many extensions thereof, radiative corrections are  required to be at least as significant as the tree-level contribution, so higher-order corrections are especially important. (2) In the Standard Model and non-supersymmetric extensions thereof a precise calculation is required to extract the parameters of the model, which when run to high energies gives information about the stability or lifetime of the potential -- which may point the way to new physics if, as appears currently, the potential is metastable. Beyond these motivations, for a generic model of new physics with boundary conditions fixed from the top down (such as  supersymmetric models) it is important to know what regions of parameter space are allowed, compatible with the measured Higgs mass. For example, a one-loop calculation may naively lead to excluding certain constrained scenarios, whereas with a two-loop calculation the Higgs mass may be large enough; this is related to the difficulty in estimating the error in the Higgs mass calculation, since at two-loop order there are new contributions from particles that have no direct couplings to the Higgs, and a simple variation of the renormalisation scale as an estimate of the error is not sufficient.

In general, there are three approaches to tackle the problem of finding the Higgs mass precisely: (i) effective potential calculations, (ii) diagrammatic calculations, (iii) renormalisation group equation methods. We shall concentrate in the following on the first two options. Calculations from the effective potential suffer from a larger uncertainty compared to diagrammatic calculations because of the missing momentum contributions. However, these are only really pronounced at one-loop level, and it is already possible to calculate the full one-loop Higgs mass inlcuding momentum dependence for generic models using \SARAH \cite{Staub:2008uz,Staub:2009bi,Staub:2010jh,Staub:2012pb,Staub:2013tta} to produce \SPheno \cite{Porod:2003um,Porod:2011nf} output  or {\tt SOFTSUSY} \cite{Allanach:2001kg}  output via {\tt FlexibleSUSY} \cite{Athron:2014yba}; explicit results have been known for some time for specific models such as the MSSM with real parameters \cite{Brignole:1991wp,Chankowski:1991md,Dabelstein:1994hb,Pierce:1996zz} and complex \cite{Pilaftsis:1998pe,Pilaftsis:1998dd,Frank:2006yh}; and for the NMSSM with real parameters \cite{Degrassi:2009yq,Staub:2010ty,Ender:2011qh} and complex \cite{Graf:2012hh}. At two loops the momentum effects are expected to be small: according to recent calculations for specific models they are comparable to the experimental uncertainty \cite{Martin:2004kr,Borowka:2014wla,Degrassi:2014pfa} and since the momentum-dependent corrections due to new physics scale at best as $m_H^2/M_{\rm New\ Physics}^2$ relative to the effective potential contribution, we expect this to be a general result. Hence effective potential calculations, with their concomitant great simplification over the diagrammatic approach, should be useful at two loop order and beyond (even if the inclusion of the momentum dependence will ultimately be necessary to reach the experimental accuracy).

In general even a two-loop calculation of the dominant contributions at zero external momentum is available for just two supersymmetric models: the MSSM \cite{Hempfling:1993qq,Carena:1995wu,Heinemeyer:1998jw,Zhang:1998bm,Heinemeyer:1998np,Heinemeyer:1999be,Espinosa:1999zm,Espinosa:2000df,Degrassi:2001yf,Brignole:2001jy,Brignole:2002bz,Martin:2002iu,Martin:2002wn,Dedes:2002dy,Dedes:2003km,Heinemeyer:2007aq,Hollik:2014bua} and partially for the next-to-minimal supersymmetric standard model (NMSSM) with real \cite{Degrassi:2009yq} and complex \cite{Muhlleitner:2014vsa} parameters. There continues to be much work in this direction and there are now some calculations of the strong (i.e. proportional to $\alpha_s$) momentum-dependent contributions for the MSSM \cite{Martin:2004kr,Borowka:2014wla,Degrassi:2014pfa}. These results have variously been made available to the community in model-specific public codes: {\tt  FeynHiggs}~\cite{Heinemeyer:1998yj}, {\tt SoftSUSY}~\cite{Allanach:2001kg}, {\tt SuSpect}~\cite{Djouadi:2002ze} and {\tt SPheno}~\cite{Porod:2003um,Porod:2011nf} for the MSSM and {\tt NMSPEC} \cite{Ellwanger:2006rn}, {\tt Next-to-Minimal SOFTSUSY} \cite{Allanach:2013kza,Allanach:2014nba}, and {\tt NMSSMCALC}  \cite{Baglio:2013iia} for the NMSSM. There are also some three-loop results, in the Standard Model \cite{Martin:2013gka,Martin:2014cxa} and the MSSM \cite{Martin:2007pg,Kant:2010tf} with the code {\tt H3m} based on \cite{Kant:2010tf}. 

The state of the art in these calculations is however somewhat suprising given that the \emph{complete generic} expressions for the two-loop effective potential, valid for a general renormalisable quantum field theory, have been available for more than ten years by the work of Martin \cite{Martin:2001vx}. These were applied to a \emph{complete} two-loop calculation of the light Higgs mass in the MSSM in the effective potential approach in Ref.~\cite{Martin:2002wn}. 
 Furthermore, generic results for the diagrammatic calculation including the momentum dependence up to leading order in gauge couplings have been available in the literature for almost as long \cite{Martin:2003it,Martin:2003qz,Martin:2005eg}. Unfortunately the results of Ref.~\cite{Martin:2002wn} suffered the so-called ``Goldstone boson catastrophe'' (recently re-explored in \cite{Martin:2014bca,Elias-Miro:2014pca}) due to the presence of tachyons in the tree-level spectrum so were numerically unstable. Perhaps due to this no public code was made available to exploit these prior to Ref.~\cite{Goodsell:2014bna} where an implementation in \SARAH/\SPheno was presented. 
Currently, the only generic two-loop results relevant for the Higgs mass calculation still not present in the literature are the all-electroweak loops and the corrections to the $Z$-boson mass relevant for determining the electroweak expectation value. These will be the subject of future work. Here we shall instead continue the process started in Ref.~\cite{Goodsell:2014bna} of making the pioneering generic results of Martin available in a public code -- which entails performing some new calculations. 

As we stated above, calculating the two-loop corrections to the Higgs mass in the effective potential approach is expected to be a good approximation. However, there is more than one way to actually perform even this calculation: either we can calculate the potential and numerically take the derivatives, as done in Refs.~\cite{Martin:2002iu} and \cite{Goodsell:2014bna}, or we can  perform the calculation diagrammatically and set the external momentum to zero. In this work we shall exploit this equivalence: we shall \emph{analytically} take the derivatives of the effective potential, producing expressions equivalent to the diagrammatic calculation and having the same structure, but with much simpler loop functions. The advantages of this over the first method are that the results are numerically stable\footnote{i.e. not subject to potential errors from ill-judged step-sizes in the numerical derivation or from parameters being too small.}; it is in principle a faster computation for more complicated models where the numerical method must make several passes to ensure stability; it can later be extended to a full diagrammatic calculation by simply changing the loop functions -- but at zero momentum the loop functions are much simpler and therefore significantly faster to evaluate. We shall therefore compute the analytic expressions for the first and second derivatives of the two-loop effective potential and implement them in \SARAH. As in Refs.~\cite{Goodsell:2014bna,Goodsell:2014pla} we shall ignore broken gauge groups, and adopt the same ans\"atze regarding the contribution of the electroweak gauge couplings to the tree-level Higgs mass matrix, to which references we refer the reader; the reasons for restricting to the so-called ``gaugeless limit'' are (a) partial circumvention of the Goldstone boson catastrophe (complete evasion in the case of the MSSM or any theory where the electroweak gauge couplings entirely determine the Higgs quartic potential); (b) significant simplification in the expressions and therefore speed in calculation; (c) the electroweak contributions are expected to be small, of the same magnitude as the momentum-dependent contributions. On the other hand, in the appendix we provide just the tadpole contributions in the case of broken gauge groups, and will return to the full expressions in future work.

The layout of the paper is as follows. In Sec.~\ref{SEC:derivatives} we explain our procedure to take the derivatives of the effective potential to extract the two-loop tadpole functions for a general theory with massless gauge bosons in a form convenient for automation. In subsection \ref{sec:Tadpoles} we summarize our results for the tadpole diagrams; we present the full set of second derivatives in appendix \ref{APP:SecondDerivatives}. The implementation of these results in \SARAH, including some technical details of the translation of the generic results into an an algorithm, is explained in Sec.~\ref{sec:sarah} before we conclude in Sec.~\ref{sec:summary}. Impatient readers interested in using our implementation of the results might want to jump directly to subsection~\ref{sec:manual}.

\section{Derivatives of the effective potential with massless vectors}
\label{SEC:derivatives}

In this section we derive the expressions for the two-loop tadpoles in a general quantum field theory with massless gauge bosons in Landau gauge. To do this, we analytically take the dervatives of the expressions in \cite{Martin:2001vx}. Writing the couplings in the notation of that paper, the theory is defined by real scalars $\phi_i$, Weyl fermions $\psi_I$ and massless gauge bosons $A_\mu^a$ where the gauge covariant derivative for the fermions and scalars are
\begin{align}
D_\mu \psi_I \equiv& \partial_\mu \psi_I+ ig A_\mu^a (T^a)_I^J \psi_J \nn\\
D_\mu \phi_i \equiv& \partial_\mu \phi_i + ig A_\mu^a \theta^a_{ij} \phi_j.
\end{align}
The structure constants $\theta^a_{ij}$ are imaginary antisymmetric matrices that obey the gauge algebra but, since we are writing in terms of real bosonic fields, for complex representations they will have twice the dimension of the equivalent generators $T^a$ (so e.g. a $U(1)$ generator is two-dimensional). We define as usual $\mathrm{tr}(\theta^a \theta^{a}) = \sum_i d(i) C(i)$ where $d(i)$ is the dimension of the representation of field $i$ and $C(i)$ the quadratic casimir, and similarly for $T^a$.

The lagrangian is then composed of the normal kinetic terms of the scalars and fermions using the above covariant derivatives supplemented by purely scalar and scalar-fermion interactions
\begin{align}
\lagr_S =& -\frac{1}{6} \lambda^{ijk} \phi_i \phi_j \phi_k - \frac{1}{24} \lambda^{ijkl} \phi_i \phi_j \phi_k \phi_l \nn\\
\lagr_{SF} =& - \frac{1}{2} y^{IJk} \psi_I \psi_J \phi_k + \cc 
\label{EQ:couplingdefinitions}\end{align}
$y$ is in general a dimensionless complex tensor with $y^{IJk}=y^{JIk}$, while $\lambda^{ijk},\lambda^{ijkl}$ are real, symmetric tensors. 

\subsection{Effective potential}
\begin{figure}[hbt]
\includegraphics[width=0.5\linewidth]{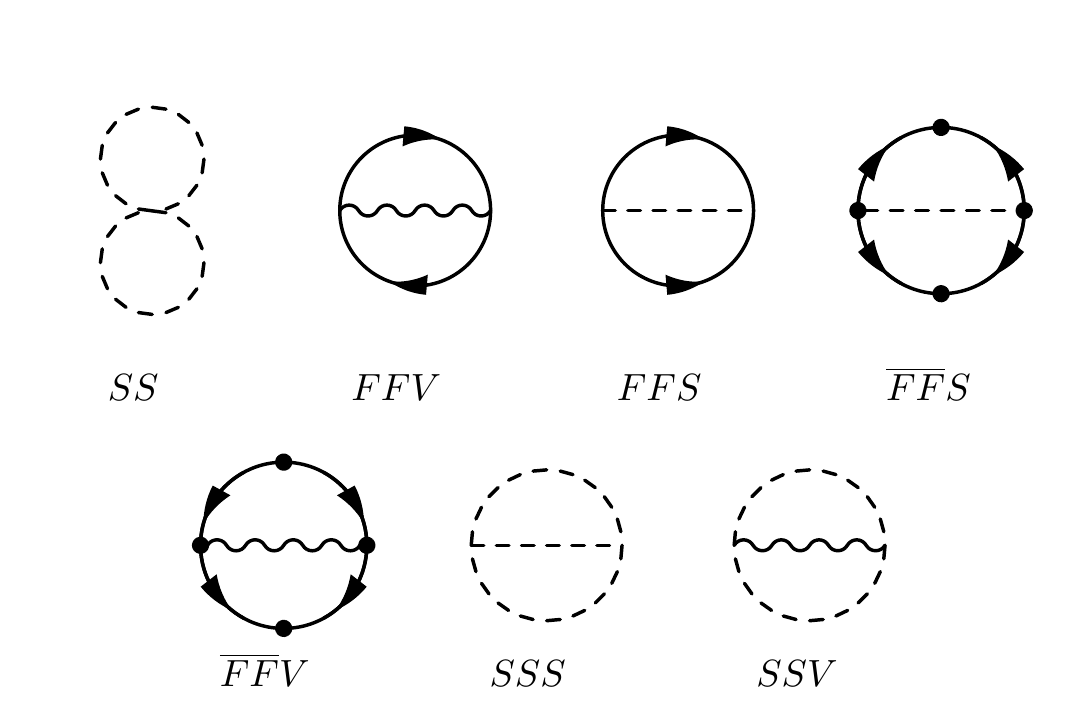}
\label{fig:EffPot}
\caption{Two-loop diagrams contributing to the effective potential in the gaugeless limit.}
\end{figure}
We can simplify the expression for the effective potential in the Landau gauge given in Ref.~\cite{Martin:2001vx} for our case; with all gauge groups unbroken some diagrams do not contribute. The non-vanishing diagrams are shown in Fig.~\ref{fig:EffPot}. \\
The contribution of each diagram to the effective potential is given by:
\begin{align}
\label{EQ:Potentialfirst}
V^{(2)}_{SSS} =& \frac{1}{12} (\lambda^{ijk})^2 f_{SSS} (m_i^2, m_j^2,m_k^2) \\
V^{(2)}_{SS} =& \frac{1}{8} \lambda^{iijj} f_{SS} (m_i^2, m_j^2) \\
V^{(2)}_{FFS} =& {1\over 2} |y^{IJk}|^2 f_{FFS} (m^2_I, m^2_J, m^2_k) \\  
V^{(2)}_{\ov{FF}S} =& {1\over 4} y^{IJk} y^{I'J'k} M^*_{II'} M^*_{JJ'} 
                             f_{\ov{FF}S} (m^2_I, m^2_J, m^2_k) + {\rm c.c.}\\
V^{(2)}_{SSV} =& {g^2\over 4} d(i) C(i) f_{SSV} (m^2_i,m^2_i, 0)\\
V^{(2)}_{FFV} =& {g^2 \over 2} d(I) C(I) f_{FFV} (m^2_I, m^2_I, 0) \\
V^{(2)}_{\ov{FF}V} =& -{g^2 \over 2} d(I) C(I) m_I^2  f_{\ov{FF}V} (m^2_I, m^2_I, 0)
\label{EQ:Potential}\end{align}
Here, $y$ and $\lambda$ are the trilinear and quartic couplings of eq.~(\ref{EQ:couplingdefinitions}), $g$ is a gauge coupling, and $M$ are fermion masses. 
The loop functions, given in terms of standard basis functions given in appendix \ref{APP:loopfunctions}, are the same for $\ov{MS}$ and $\ov{DR}^\prime$ in the case of 
\begin{align}
f_{SSS}= & - I(x,y,z) \nn \\
f_{SS}= & J(x,y) \nn\\ 
f_{FFS}= & J(x,y) - J(x,z) - J(y,z) + (x + y - z) I(x,y,z) \nn\\
f_{\ov{FF}S} =& 2 I(x,y,z) \nn\\
f_{SSV} = & (x+y)^2 + 3 (x+y)I(x,y,0) + 3J(x,y) - 2xJ(x) - 2yJ(y)
\end{align}
but differ for those with vectors and fermions:
\begin{equation}
\begin{array}{|c|c|c|} \hline 
 & \ov{MS} & \ov{DR}^\prime \\ \hline
f_{FFV} & 0 & - (x+y)^2 + 2x J(x) + 2y J(y)\\
f_{\ov{FF}V} & 6I (x,y,0) + 2(x+y) - 4 J(x) - 4 J(y) & 6I (x,y,0)\\ \hline
\end{array}
\end{equation}
All of these functions are symmetric on the substitution of their first two indices, but may not be so with the third. In fact, we can then combine the vector-fermion diagrams to give
\begin{align}
V^{(2)}_{FFV}+V^{(2)}_{\ov{FF}V} \equiv {g^2 \over 2} d(I) C(I) F_{FV} (m_I^2)
\end{align}
where
\begin{align}
 F_{FV} (x) \equiv&  -4 x^2 + 4 x J(x) - 6 x I(x,x,0) + \delta_{\rm{MS}} 4x J(x)
\end{align}
where $ \delta_{\rm{MS}} $ is zero for $\ov{\rm{DR}}^\prime$ and one for $\ov{\rm{MS}}$.

\subsection{Derivatives of the potential}

Here we shall analytically take the derivatives of the potential. In a generic model, we may want the derivatives of the Higgs potential in terms of some unrotated fields, i.e. in a basis where the mass matrix is not diagonal; let us say that we start with such a case. We write the tree-level potential in terms of some expectation values $\hat{v}_i$ and the associated real fluctuations $S_i^0$ as
\begin{align}
V^{\mathrm{scalar\ tree}} = V_0 (\hat{v}_i) + \frac{1}{2} \hat{m}_{ij}^2 S_i^0 S_j^0 + \frac{1}{6} \hat{\lambda}_0^{ijk} S_i^0 S_j^0 S_k^0 + \frac{1}{24} \hat{\lambda}^{ijkl}_0 S_i^0 S_j^0 S_k^0 S_l^0.
\end{align}
Then there is a tree-level rotation
\begin{equation}
S_i^0 = R^0_{ij} S_j
\end{equation}
to diagonalise the mass matrix; we then obtain
\begin{align}
V^{\mathrm{scalar\ tree}} = V_0 + \frac{1}{2} m_{i}^2 S_i S_i + \frac{1}{6} \lambda^{ijk} S_i S_j S_k + \frac{1}{24} \lambda^{ijkl} S_i S_j S_k S_l.
\end{align}
We can write $\phi_i = v_i + S_i$ and work with the couplings of eq.~(\ref{EQ:couplingdefinitions}). We then need the quantities which enter in the effective potential calculation, which are masses and couplings depending on $\{S_i\}$.\\
In general, we have
\begin{align}
m_{ij}^2 (S) =& \frac{\partial^2}{\partial S_i \partial S_j} V \nn\\
=& m_{ i}^2 \delta_{ij}+ \lambda^{ijk} S_k + \frac{1}{2} \lambda^{ijkl} S_k S_l\nn\\
\lambda^{ijk} (S) =& \frac{\partial^3}{\partial S_i \partial S_j \partial S_k} V \nn\\
&= \lambda^{ijk} + \lambda^{ijkl}  S_l \nn\\
\lambda^{ijkl} (S) =& \lambda^{ijkl}
\end{align}
with the shorthand notation $S\equiv\{S_i\}$.
Hence we can write
\begin{align}
\frac{\partial}{\partial S_r} \lambda^{ijkl} (S) =& 0 \nn\\
\frac{\partial}{\partial S_r} \lambda^{ijk} (S) =& \lambda^{ijkr} (S) \nn\\
\frac{\partial}{\partial S_r} m_{ij}^2 (S) =& \lambda^{ijr} +  \lambda^{ijkr} S_k .
\end{align}
Similarly for the fermions we have
\begin{align}
m_I^2 \delta_J^I =& M^{I I'}M_{JI'}^* \nn\\
\rightarrow \frac{\partial}{\partial S_r} M^{I I'}M_{JI'}^* =& y^{II'r} M_{JI'}^* + M^{I I'} y_{JI'r} \nn\\
 \frac{\partial}{\partial S_r} y^{IJs} =& 0.
\end{align}

However, for the purposes of the effective potential, we then require a further diagonalisation for $m_{ij}^2 (S) $: we put 
\begin{equation}
S_i = R_{ij} (S)\, S_j^\prime.
\end{equation}
We name the couplings in this basis as $\tilde{m}_i(S), \tilde{\lambda}^{ijk}_S(S),\tilde{\lambda}^{ijkl}(S) $.
We  then express the effective potential by inserting the couplings and masses in the basis $\{S_i^\prime\}$ into the formulae of eqs.~(\ref{EQ:Potentialfirst})--(\ref{EQ:Potential}). However, to take the derivatives we rewrite the expressions in terms of the basis $\{S_i\}$, and use the trick (with ${\bf m^2}\equiv(m_{ij}^2)$)
\begin{align}
 \frac{\partial}{\partial S_r} \left( \frac{1}{q^2 + {\bf m^2}} \right)_{ij} =& -\left(\frac{1}{q^2 + {\bf m^2}}\right)_{ik} \frac{\partial m_{k k'}^2}{\partial S_r} \left(\frac{1}{q^2 + {\bf m^2}}\right)_{k'j} \,.
\end{align}
For similar expressions we write by abuse of notation (using $C\equiv 16\pi^2 \mu^{2\epsilon}(2\pi)^{-d}$ \cite{Martin:2003qz})
\begin{align}
  \mathbf{J}(m^2_{ik},m_{jl}^2) &\equiv C^2\int d^dq d^dk \left( \frac{1}{q^2 + {\bf m^2}} \right)_{ik}  \left( \frac{1}{k^2 + {\bf m^2}} \right)_{jl}.
\end{align}
For fermion propagators we can write
\begin{align}
\frac{M_{II'}}{q^2 + m_{I}^2} \rightarrow M_{IJ} \frac{1}{q^2 + M^{JK}M_{KI'}} =  \frac{1}{q^2 + M_{IJ}M^{JK}} M_{KI'}.
\end{align}
Let us demonstrate our method on a brief example:
\begin{align}
\frac{\partial}{\partial S_p^0 }\frac{1}{8} \tilde{\lambda}^{iijj} f_{SS} (\tilde{m}_i^2, \tilde{m}_j^2) =&  R_{rp}^0\frac{\partial}{\partial S_r} \frac{1}{8} \tilde{\lambda}^{iijj} f_{SS} (\tilde{m}_i^2, \tilde{m}_j^2) \nn\\
=& R_{rp}^0\frac{\partial}{\partial S_r} \frac{1}{8} \lambda^{ik jl} f_{SS} (m_{ik}^2, m_{jl}^2) \nn\\
=& R_{rp}^0 \frac{1}{4}\lambda^{ik jl} (S) f^{(1,0)}_{SS} (m_{im}^2,m_{nk}^2; m_{jl}^2) \frac{\partial}{\partial S_r}( m_{mn}^2) \nn\\
=& R_{rp}^0\frac{1}{4}\lambda^{ik jl} (S) f^{(1,0)}_{SS} (m_{im}^2,m_{nk}^2; m_{jl}^2) \lambda^{mn r} (S) \nn\\
\underset{S\rightarrow 0}{\longrightarrow}& R_{rp}^0\frac{1}{4}\lambda^{ik jj} \lambda^{i k r} f^{(1,0)}_{SS} (m_i^2, m_k^2;m_j^2 ).
\end{align}
Recall that $f_{SS} (x,y) \equiv J(x,y)$ where $J$ is the finite loop function and ${\bf J}(x)=J(x)-\frac x\epsilon$, see Eq.~(\ref{eq:oneloopfunctionsAJ}); here we have defined 
\begin{align}
f^{(1,0)}_{SS} (x, y; z)\equiv&  -C^2 \int d^d k d^d q \frac{1}{k^2 + x} \frac{1}{k^2 + y} \frac{1}{q^2 + z} + \frac{C}{\epsilon}\bigg({\bf J}(z) - C z \int d^d k \frac{1}{k^2 + x} \frac{1}{k^2 + y}  \bigg) + \frac{z}{\epsilon^2} \nn\\
=& -C^2 \bigg( \int d^d k \frac{1}{k^2 + x} \frac{1}{k^2 + y} \bigg) \bigg(\int d^d q  \frac{1}{q^2 + z} +\frac{z}{\epsilon} \bigg) + \frac{C}{\epsilon}{\bf J}(z) + \frac{z}{\epsilon^2}\nn\\
=& \frac{1}{x-y} \bigg( {\bf J}(x) - {\bf J}(y) \bigg) J(z) + \frac{1}{\epsilon} J(z) \nn\\
=&  \frac{1}{x-y} \bigg( J(x) - J(y) \bigg) J(z) \nn\\
=& - B_0 (x,y) J(z).
\end{align}
Note that the $R_{ra}^0$  are not functions of $S_i$ and so do not present complications if we want to take futher derivatives. \\

We can similarly take the derivatives of all the remaining loop functions; these are derived from the basis:
\begin{align}
\frac{\partial}{\partial S_r} J(m_i^2,m_j^2) \rightarrow & -B_0 (m_i^2,m_k^2) J(m_j^2)  \frac{\partial m_{ik}^2}{\partial S_r} - J(m_i^2) B_0 (m_j^2,m_k^2) \frac{\partial m_{jk}^2}{\partial S_r} \nn\\
\frac{\partial}{\partial S_r} B_0 (m_i^2,m_j^2) \rightarrow& - C_0 (m_i^2,m_k^2,m_j^2) ( \frac{\partial m_{ik}^2}{\partial S_r} +  \frac{\partial m_{jk}^2}{\partial S_r}) \nn\\
\frac{\partial}{\partial S_r} I(m_i^2,m_j^2,m_k^2) \rightarrow& - U_0 (m_i^2,m_l^2,m_j^2,m_k^2)  \frac{\partial m_{il}^2}{\partial S_r} - U_0 (m_j^2,m_l^2,m_i^2,m_k^2)  \frac{\partial m_{jl}^2}{\partial S_r} \nn\\
&- U_0 (m_k^2,m_l^2,m_i^2,m_j^2)  \frac{\partial m_{kl}^2}{\partial S_r}
\end{align}
and
\begin{align}
\frac{\partial}{\partial S_r} m_i^2 I(m_i^2,m_j^2,m_k^2) \rightarrow& \frac{\partial m_{il}^2}{\partial S_r} I(m_l^2,m_j^2,m_k^2) - m_i^2  \frac{\partial m_{il}^2}{\partial S_r} U_0 ( m_l^2,m_i^2,m_j^2,m_k^2) \nn\\
&-\frac{\partial m_{jl}^2}{\partial S_r}m_i^2 U_0 (m_l^2,m_j^2,m_k^2,m_i^2) - \frac{\partial m_{kl}^2}{\partial S_r}m_i^2 U_0 (m_l^2,m_k^2,m_i^2,m_j^2) .
\end{align}

The derivative of a typical term in the effective potential will have the form
\begin{align}
\frac{\partial}{\partial S_r} A^{ijk} A^{ijk} f_\alpha(x_i,y_j,z_k) =& 2 f_\alpha(x_i,y_j,z_k) A^{ijk}  \frac{\partial}{\partial S_r} (A^{ijk})  \nn\\
&+ \bigg\{  A^{ijk} A^{i'jk} \frac{\partial m^2_{ii'}}{\partial S_r} f_\alpha^{(1,0,0)}(x_i,x_{i'};y_j,z_k) + (x\leftrightarrow y) + (x\leftrightarrow z)\bigg\}.
\end{align}
where, generalising the above, it is straightforward to show that for a generic function appearing in the effective potential $f_\alpha$ composed of polynomials (even containing monomials with negative exponents) multiplying the loop functions above, we can write
\begin{align}
f_\alpha^{(1,0,0)} (x,u; y,z) \equiv& \frac{f_\alpha (x,y,z) - f_\alpha(u,y,z)}{x-u}
\end{align}
and similarly for permutation of the indices. On the other hand, this explicit expression is often inconvenient in practice due to the need to carefully take the smooth limit when $x=u$; it is instead more practical to rewrite the right hand side in terms of our basis of loop functions multiplied by suitable polynomials. In the following we present explicit expressions for the first derivatives (and, in the appendix,  the second derivatives) which have been appropriately simplified to remove the apparent singularities.

\subsection{First derivatives}
\label{sec:Tadpoles}

\begin{figure}[hbt]
\includegraphics[width=0.4\linewidth]{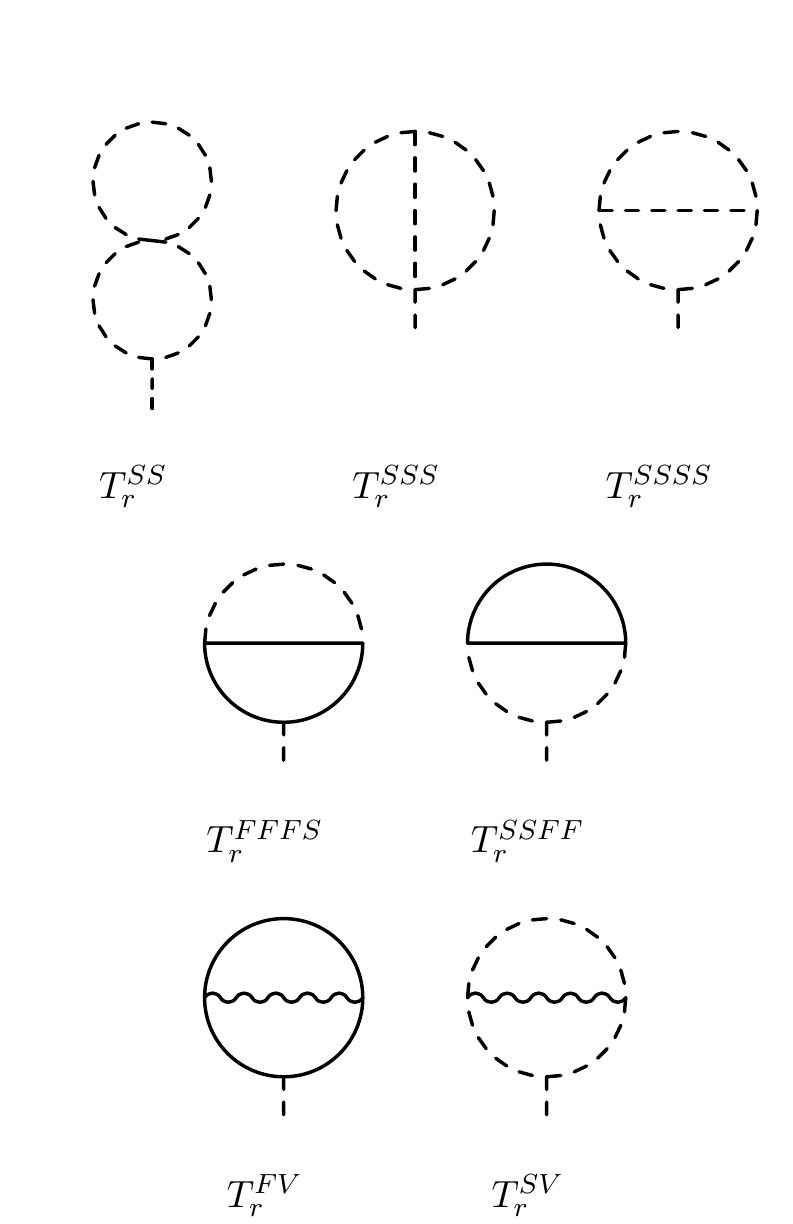}
\label{fig:Tad2L}
\caption{Tadpole diagrams at the two-loop level which don't vanish in the gaugeless limit. }
\end{figure}

Here we gather the set of two-loop tadpole diagrams. There are only three topologies, two of which only apply to the all-scalar case. All generic possible diagrams are given in Fig.~\ref{fig:Tad2L}. Here we present our results for the first derivatives in the basis $\{S_i\}$, so without the rotation matrices $R$; the full tadpole in the original basis is then given by
\begin{align}
\frac{\partial V^{(2)}}{\partial S_p^0} = R_{rp}^0 [T_{S} + T_{SSFF}+ T_{FFFS}+T_{SV}+T_{FV}]
\label{EQ:FullTadpoleGaugeless}\end{align}
with the tadpoles on the right-hand side to be defined below.

\subsubsection{All scalars}
We start with the purely scalar diagrams which are in the first row of Fig.~\ref{fig:Tad2L}. The entire contribution is given by
\begin{equation}
T_{S}  = T_{SS} +  T_{SSS} + T_{SSSS}
\end{equation}
with 
\begin{align}
T_{SS} =& \frac{1}{4}\lambda^{ik jj} \lambda^{i k r} f^{(1,0)}_{SS} (m_i^2, m_k^2;m_j^2 ) \\
T_{SSS} =&\frac{1}{6} \lambda^{rijk} \lambda^{i jk} f_{SSS} (m_i^2, m_j^2,m_k^2) \\
T_{SSSS} = & \frac{1}{4} \lambda^{rii'} \lambda^{ijk} \lambda^{i'jk} f_{SSS}^{(1,0,0)} (m_{i}^2,m_{i'}^2; m_{j}^2,m_{k}^2) .
\end{align}
The new loop functions are defined as 
\begin{align}
f^{(1,0)}_{SS} (x,y;z ) \equiv&- B_0 (x,y) J(z) \nn\\
f_{SSS}^{(1,0,0)} (x,y;u,v ) \equiv& U_0 (x,y,u,v).
\end{align}
Note that $f^{(1,0)}_{SS}$ corresponds to $X_{SSS}$, $f_{SSS}$ corresponds to $S_{SSS}$, and $f^{(1,0,0)}_{SSS}$ to $W_{SSSS}$ of Ref.~\cite{Martin:2003it} in the limit of zero external momentum.

\subsubsection{Scalars and fermions}
We have, first, the diagrams with two scalar propagators:
\begin{align}
T_{SSFF}=& \frac{1}{2}y^{IJk} y_{IJl} f_{FFS}^{0,0,1} (m^2_I, m^2_J; m_k^2,m_l^2) \lambda^{klr} \nn\\
& - \bigg[\frac{1}{2} y^{IJk} y^{I'J'k} M^*_{II'} M^*_{JJ'} \lambda^{klr} U_0(m_l^2,m^2_k,m^2_I, m^2_J) + c.c.\bigg].
\end{align}
Then we turn to one scalar and three fermion propagators:
\begin{align}
T_{FFFS}=& 2\mathrm{Re} \big[ y^{IJr} y_{IKm} y^{KLm}  M^*_{JL} \big] T_{F\ov{F}FS} (m_I^2, m_J^2,m_K^2,m_m^2) \nn\\
& + 2\mathrm{Re} \big[ y_{IJr} y^{IKm} y^{JLm} M^*_{KL} \big] T_{FF\ov{F}S} (m_I^2, m_J^2,m_K^2,m_m^2) \nn\\ 
&- 2\mathrm{Re} \big[ y^{IJr} y^{KLm} y^{MNm} M^*_{IK}M^*_{JM} M^*_{LN}\big] T_{\ov{F}\ov{F}\ov{F}S} (m_I^2, m_J^2,m_L^2,m_m^2).
\end{align}
Here we have defined
\begin{align}
f_{FFS}^{1,0,0} (m_{I'}^2,m^2_I, m^2_J; m_k^2) \equiv & - B_0 (m_{I'}^2,m^2_I) J(m_J^2) + B_0 (m_{I'}^2,m^2_I) J(m_k^2) + I( m_{I'}^2, m_J^2,m_k^2)\nn\\
&- (m_I^2 + m_J^2 - m_k^2) U_0(m_{I'}^2,m^2_I, m^2_J, m_k^2), \nn\\
f_{FFS}^{0,0,1} (m^2_I, m^2_J; m_k^2, m_l^2) \equiv &  B_0 (m_l^2,m^2_k) J(m_I^2) + B_0 (m_l^2,m^2_k) J(m_J^2) - I( m_{I}^2, m_J^2,m_l^2)\nn\\
&- (m_I^2 + m_J^2 - m_k^2) U_0(m_l^2,m_k^2,m^2_I, m^2_J) \nn\\
T_{\ov{F}\ov{F}\ov{F}S} (m_I^2, m_J^2,m_L^2,m_m^2) \equiv& U_0( m_I^2, m_J^2,m_L^2, m_m^2). \nn\\
 T_{F\ov{F}FS} (m_I^2, m_J^2,m_K^2,m_m^2) \equiv& f_{FFS}^{1,0,0} (m_I^2, m_J^2,m_K^2;m_m^2) \nn\\
 T_{FF\ov{F}S} (m_I^2, m_J^2,m_K^2,m_m^2) \equiv&  I(m_I^2,m_K^2,m_m^2) - m_I^2 U_0(m_I^2, m_J^2,m_K^2,m_m^2).
\end{align}

\subsubsection{Diagrams with vectors}
Finally, the two generic diagrams involving vectors are given by
\begin{align}
T_{SV}=&   \frac{g^2}{2 } d(i) C(i) \lambda^{iir}\bigg(3I(0,m_i^2,m_i^2) - J(m_i^2) +2 m_i^2 \bigg) \\
=&  \frac{g^2}{2 } d(i) C(i) \lambda^{iir} m_i^2 \bigg[ -12 + 11 \log m_i^2/Q^2 - 3 \log^2 m_i^2/Q^2) \bigg] .\label{EQ:TadpoleScalarVector}\nn\\
T_{FV}=&  g^2  d(I) C(I) \mathrm{Re}( M_{I I'} y^{II'r}  ) 4 \bigg(-3I(0,m_I^2,m_I^2) + 5 J(m_I^2) -4 m_I^2 + \delta_{\ov{\mathrm{MS}}} \big[ 2 J(m_I^2) + m_I^2\big] \bigg)\\
=&  g^2  d(I) C(I) \mathrm{Re}( M_{I I'} y^{II'r}  ) 4m_I^2 \bigg[ 6 - 7 \log m_I^2/Q^2 +3 \log^2 m_I^2/Q^2  + \delta_{\ov{\mathrm{MS}}} \big[ 2 \log m_I^2/Q^2 -  1\big]\bigg]. \nn
\end{align}

\subsection{Second derivatives}

\begin{figure}[hbt]
\includegraphics[width=0.66\linewidth]{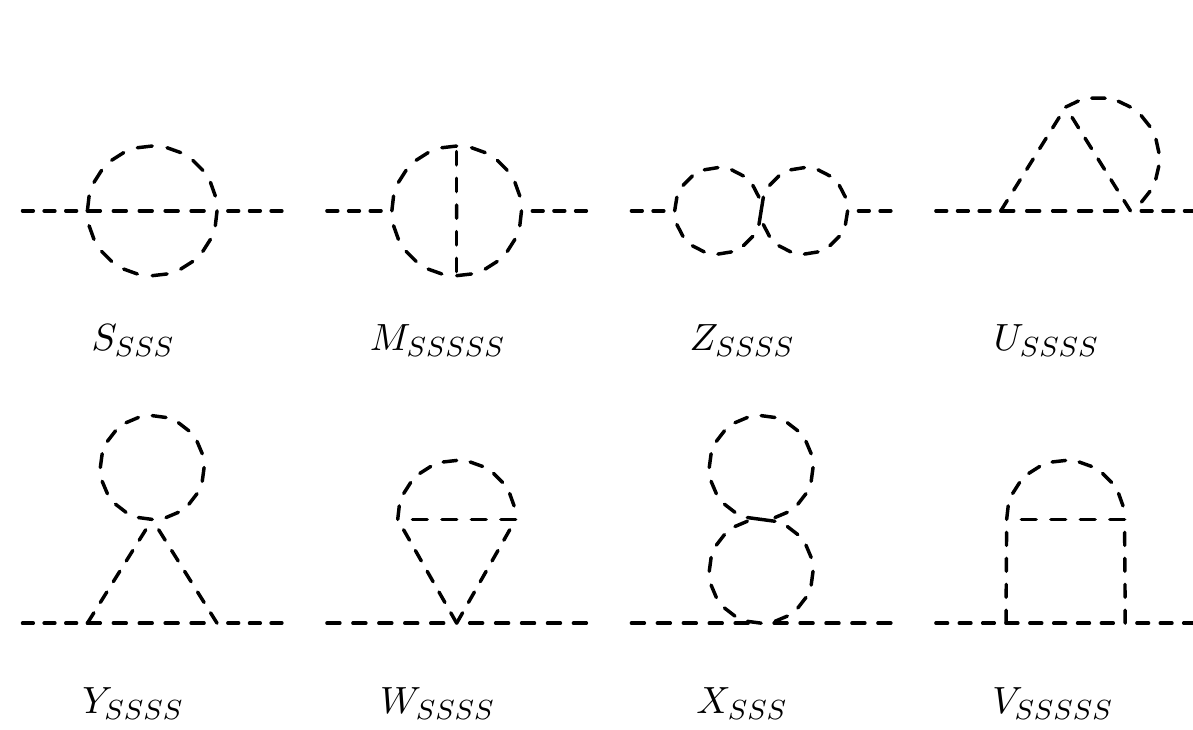}
\caption{Two-loop self-energy diagrams involving only scalars. }
\label{fig:SelfScalars}\end{figure}

\begin{figure}[hbt]
\includegraphics[width=0.5\linewidth]{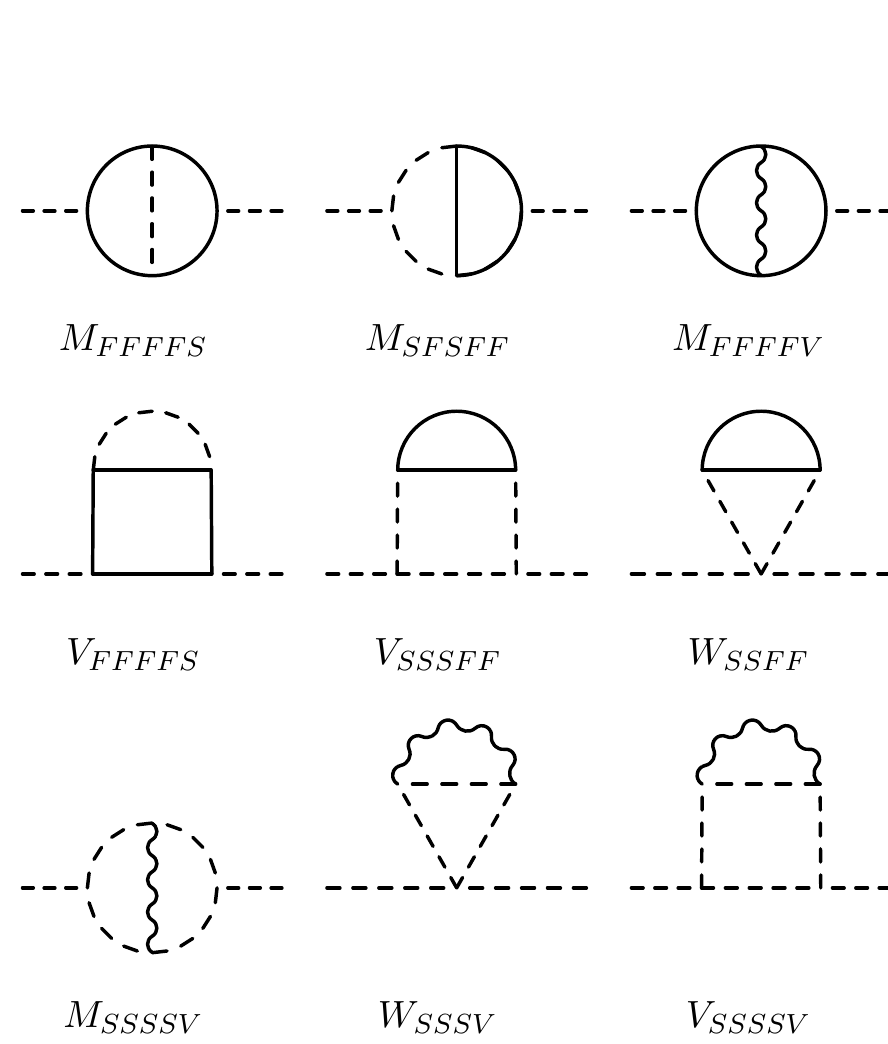}
\caption{Remaining two-loop self-energy diagrams which do not vanish in the gaugeless limit. }
\label{fig:SelfMixed}\end{figure}

To find the second derivatives of the potential we can apply the same technique. However, in principle, we can simply use the results of \cite{Martin:2003it}, which computed (diagrammatically) the two-loop scalar self-energies at leading order in gauge couplings. Since we want the self-energies for neutral scalars, this comprises all of the contributions, and if we want the results in the effective potential approach we can simply set the external momenta to zero (and, for this work, the masses of the gauge bosons to zero). In fact, for the majority of the diagrams, these yield the same result. However, in a few cases we find that by taking the derivatives of the potential we find simpler results (which are of course entirely equivalent).
The full result is given by
\begin{align}
\frac{\partial V^{(2)}}{\partial S_p^0\partial S_q^0} =& R_{ip}^0 R_{jq}^0 \Pi_{ij} (0) \nn\\
\equiv& R_{ip}^0 R_{jq}^0 \bigg[ \Pi^{S}_{ij} + \Pi^{SF(W)}_{ij} + \Pi^{SF_4 (M)}_{ij} + \Pi^{S_2 F_3 (M)}_{ij} + \Pi^{S_3F_2 (V)}_{ij} + \Pi^{SF_4 (V)}_{ij} + \Pi^{SV}_{ij} + \Pi_{ij}^{FV}\bigg]
\end{align}
where the superscripts correspond to the numbers of scalars, fermions and vectors with types of topology listed for diagrams $(M)$ and $(V)$.
We give the complete set of relevant expressions in  appendix \ref{APP:SecondDerivatives}; the corresponding diagrams are shown in Figs.~\ref{fig:SelfScalars} and \ref{fig:SelfMixed}. The expressions for $ \Pi^{SF_4 (V)}_{ij}, \Pi^{SV}_{ij}$ and $\Pi_{ij}^{FV} $  exhibit particular simplifications in our approach.

\label{sec:selfenergies}

\section{Implementation in \SARAH}
\label{sec:sarah}

We have implemented the new routines in \SARAH. By including the first and second derivatives of the effective potential using the analytic expressions here, rather than numerically taking the derivatives of the couplings and masses as performed in the previous version \cite{Goodsell:2014bna}, we can guarantee greater numerical stability, accuracy and speed improvements. Moreover, this approach allows a straightforward upgrade to the \emph{pole} mass calculation by simply changing the loop functions called to those defined in Ref.~\cite{Martin:2003it} based on loop functions implemented in {\tt TSIL} \cite{Martin:2005qm}, which will be made possible in a future version. 

\subsection{Method}
For any given supersymmetric\footnote{A two-loop Higgs mass calculation in non-SUSY models will become available in future releases} model {\tt \$MODEL}, once the user has  specified the particle content and their symmetries, \SARAH calculates all of the vertices and masses. It then writes a \Fortran code (placed in the suggestively named {\tt 2LPole\_\$MODEL.f90}) %$
which implements our expressions, linking to a static \Fortran code (named {\tt 2LPoleFunctions.f90}) of the basis functions for the generic first and second derivatives of the effective potential defined in section \ref{SEC:derivatives} and the appendix. These two pieces of \Fortran code are called by \SPheno during the calculation of the loop corrections to the Higgs mass. Here we shall give a few details of how \SARAH writes {\tt 2LPole\_\$MODEL.f90}. 
The overall algorithm is to 
\begin{enumerate}
\item Generate masses and couplings for all relevant particles in the gaugeless limit.
\item Populate and classify all tadpole topologies according to particle content.
\item For each tadpole topology, pass the set of diagrams along with information specifying the symmetries to a generic writer function.
\item Rotate the total tadpole vector by the Higgs rotation matrix to the non-diagonal basis, c.f. eq.~(\ref{EQ:FullTadpoleGaugeless}).
\item Populate and classify all second derivative topologies according to particle content.
\item For each mass topology, pass the set of diagrams along with information specifying the symmetries to a generic writer function.
\item Rotate the mass matrix to the non-diagonal basis (as with the tadpoles).
\end{enumerate}
The writer function is actually identical for tadpoles and mass diagrams with a switch to adjust the number of Higgses. It cycles through the list of diagrams and applies the following process for each:
\begin{enumerate}[(a)]
\item Determine symmetry factor of diagram due to permutations.
\item Determine the colour factor; for diagrams with a gluon propagator this is simply $d(I) C(I)$ whereas otherwise we must trace over colour indices of the vertices. In principle, for four-point vertices there can be two colour structures for the vertex which superficially leads to more than one colour factor for such diagrams. However, as we can simply see by inspecting the expressions in the appendix,  or by considering that the colour factors have to be inherited from a corresponding vacuum diagram (since differentiating with respect to neutral Higgs fields cannot introduce any additional colour factor), for diagrams with a four-point vertex consisting of four coloured fields the colour factor is given by a trace over the indices in pairs. Hence such four-point vertices are saved with the colour factor of the pairs of indices traced over. To be more explicit, such vertices can only contribute if they come from differentiating $V_{SS}^{(2)}$ which contains the coupling $\lambda^{iijj}$. We then must simply take care that the indices of the vertices correspond correctly to the indices that are traced over.
\item Write a nested set of loops to sum the diagram over the generations of all particles and, for the inner loop, the external Higgs legs, since the most computationally demanding aspect is evaluating the loop functions and this can be evaluated before calling the Higgs loop -- indeed, we also check that the coupling multiplying it is non-zero first too.
\end{enumerate}

There are subtleties in translating our results into a form usable by \SARAH, both stemming from the fact that in the \SPheno code the couplings are stored in terms of either real or complex scalars, and four-component spinor fermions, while, since our results are based on those of Refs.~\cite{Martin:2001vx} and \cite{Martin:2003it} and for economy we use real scalars and two-component fermions. The translation between the two bases as required by \SARAH and \SPheno is largely as described in Ref.~\cite{Goodsell:2014bna}, however here we have the additional complication for fermions of translating chains of couplings and masses such as 
$$C_1=\mathrm{Re}( y^{IJr} y^{KLm} y^{MNm} M^*_{IK}M^*_{JM} M^*_{LN}). $$
In \SARAH, the interactions of Weyl spinors $\psi$ are derived from the corresponding Dirac spinor interactions $\Psi$ as 
\begin{align}
\mathrm{Vertex}  = i \frac{\delta \lagr}{\delta \phi_r \delta \bar \Psi_I \delta \Psi_J} = i \left(\frac{\delta \lagr}{\delta \phi_r \delta \psi^R_I \delta \psi^L_J} P_L + \frac{\delta \lagr}{\delta \phi_r \delta \psi^{L*}_I \delta \psi^{R*}_J} P_R\right)  \equiv i (c_L P_L + c_R P_R) ,
\end{align}
with $\psi^Y_X = P_Y \Psi_X$ ($Y=L,R$; $X=I,J$) and polarization operators $P_L$, $P_R$;
and so $c_{L,R} \leftrightarrow -y^{IJr}, c_{L,R}^* \leftrightarrow -y_{IJr}$. When we have complex scalars $\Phi$, we should write
\begin{align}
\mathcal{L} \supset&  \Phi^m (\ov{\Psi}_I (c_L (I,J,m) P_L + c_R (I,J,m) P_R) \Psi_J  +  \Phi_m (\ov{\Psi}_I (c_L (I,J,\ov{m}) P_L + c_R (I,J,\ov{m})P_R) \Psi_J \nn\\
=& \Phi^m (\ov{\Psi}_I (c_L (I,J,m) P_L + c_R (I,J,m) P_R) \Psi_J  +  \Phi_m (\ov{\Psi}_J (c_R^* (I,J,m) P_L + c_L^* (I,J,m) P_R) \Psi_J
\end{align}
and so $c_L(I,J,\ov{m}) = c_R^*(J,I,m)$. For a given topology, \SARAH populates the diagrams using Dirac propagators (i.e. links fermions with its conjugate) and so for the coupling $C_1$ above we will find sets of particles
\begin{align}
\{\ov{\Psi}_I,\Psi_J,\phi^r\},\{\ov{\Psi}_N,\Psi_I,\phi^m\},\{\ov{\Psi}_J,\Psi_N,\ov{\phi}_m\}. 
\end{align}
Suppose each of the fermions is a Dirac spinor with Weyl spinors $\psi^I_{L,R}$ etc, then to construct coupling $C$ above we must sum over the left and right-handed Weyl fermions (which have opposite representations of all gauge groups) and thus (noting that \SPheno always internally stores the fermion masses  as real positive definite)
\begin{align}
C_1 = \mathrm{Re} \bigg[&y^{\psi^I_L,\psi^J_R,r}y^{\psi^N_L,\psi^I_R,m}y^{\psi^J_L,\psi^N_R,\ov{m}} + y^{\psi^I_R,\psi^J_L,r}y^{\psi^N_R,\psi^I_L,\ov{m}}y^{\psi^J_R,\psi^N_L,m}\bigg] M_I M_J M_L \nn\\
=& -\mathrm{Re} \bigg[c_R^* (I,J,r) c_R^* (N,I,\ov{m}) c_R^*(J,N,m) + c_L(I,J,r) c_L(N,I,\ov{m}) c_L(J,N,m) \bigg] M_I M_J M_N \nn\\
=& -\mathrm{Re} \bigg[c_R (I,J,r) c_R (N,I,\ov{m}) c_R(J,N,m) + c_L(I,J,r) c_L(N,I,\ov{m}) c_L(J,N,m) \bigg] M_I M_J M_L.
\end{align}
On the other hand, if the fermions are all Majorana then $\psi^I_L = \psi_R^I$ and we therefore only have half of this sum, so we include an extra factor of $1/2$.
If we consider another example
\begin{align}
C_2 =& \mathrm{Re}( y^{IJr} y_{IKm} y^{KNm} M_{JN}^* )
\end{align}
in \SARAH we would generate the set of particles 
\begin{align}
\{\ov{\Psi}_I,\Psi_J,\phi^r\},\{\ov{\Psi}_K,\Psi_I,\ov{\phi}_m\},\{\ov{\Psi}_J,\Psi_K,\phi^m\}. 
\end{align}
and thus
\begin{align}
C_2 \rightarrow& \mathrm{Re} \bigg[y^{\psi^I_L,\psi^J_R,r}y_{\psi^I_L,\psi^K_R,\ov{m}}y^{\psi^K_R,\psi^N_L,m}+y^{\psi^I_R,\psi^J_L,r}y_{\psi^I_R,\psi^K_L}^my^{\psi^K_L,\psi^N_R}_{\ov{m}} \bigg]  M_J  \nn\\
\rightarrow& -\mathrm{Re} \bigg[c_R^* (I,J,r) c_L^* (K,I,m) c_R^*(J,K,m) +  c_L (I,J,r) c_R (K,I,m) c_L(J,K,m)\bigg] M_J \nn\\
=&-\mathrm{Re} \bigg[c_R (I,J,r) c_L (K,I,m) c_R(J,K,m) +  c_L (I,J,r) c_R (K,I,m) c_L(J,K,m)\bigg] M_J .
\end{align}
 The above show that it is straightforward to translate the two-component results into expressions in \SARAH 
\begin{align}
\mathrm{Re}\bigg[& \big[ \prod_{i=1}^m y^{I_i J_i s_i}  \prod_{j=1}^n y_{I_j J_j s_j} \big] \big[ \prod_{k=1}^p M_{I_k J_k} \big] \bigg] \rightarrow\nn\\
& \left(\frac{1}{2} \right)^M (-1)^{m+n}  \mathrm{Re}\bigg[ \prod_{i=1}^mc_L(I_i,J_i,s_i) \prod_{j=1}^nc_R(I_j,J_j,s_j) +  \prod_{i=1}^mc_R(I_i,J_i,s_i)\prod_{j=1}^nc_L(I_j,J_j,s_j)\bigg] \big[ \prod_{k=1}^p M_{I_k} \big] 
\end{align}
where $M=1$ for Majorana fermions and zero otherwise.

A final point regarding the translation into a basis of real and complex scalars is that the new routines assume that there is a unique way of constructing a gauge- and global symmetry-invariant coupling $\lambda^{ijk}$ from complex scalars other than the complex conjugate of the whole coupling; i.e. if $\lambda^{ijk}$ is permitted for given complex $i,j,k$ then $\lambda^{ij}_k$ is not. This is evidently true -- if both are permitted then we can generate a holomorphic mass term at one loop which violates the premise. However, it is important in that we cannot write gauge singlets as complex scalars if they have couplings violating the above condition, no matter how small the couplings -- for example for sneutrinos in see-saw models.

\subsection{How to use the new routines}
\label{sec:manual}

To study a model with \SARAH the general procedure is as follows: the user should download and run \SARAH with the demanded model. \SARAH derives all analytical expressions for mass matrices, vertices, renormalisation group equations as well as loop corrections and exports this information into \Fortran source code. The \Fortran source code is compiled together with \SPheno and all numerical 
calculations are then performed by the new \SPheno module. This includes a calculation of the entire mass spectrum, branching ratios as well as flavour and other precision observables \cite{Porod:2014xia}. For the mass spectrum all one-loop corrections to any particle are included in a diagrammatic way \cite{Staub:2010ty}. For a supersymmetric model there are now in addition three options to get two-loop corrections in the Higgs sector. The first two are based on the effective potential approach presented in Ref.~\cite{Goodsell:2014bna} while the new routines are called by the (now default) third option. 

A step-by-step description to obtain a spectrum generator for an arbitrary model {\tt \$MODEL} implemented in \SARAH reads as follows:
\begin{enumerate}
\item Download the most recent \SARAH and \SPheno versions into a directory {\tt \$PATH}. Both packages are located at {\tt HepForge}:
\begin{lstlisting}
http://sarah.hepforge.org/
http://spheno.hepforge.org/ 
\end{lstlisting}
\item Enter the directory and extract both codes 
\begin{lstlisting}
> cd $PATH
> tar -xf SARAH-4.5.0.tar.gz
> tar -xf SPheno-3.3.3.tar.gz 
\end{lstlisting}
\item Start \Mathematica, load \SARAH, run {\tt \$MODEL}, and generated the \SPheno output
\begin{lstlisting}
<< $PATH/SARAH-4.5.0/SARAH.m;
Start["$MODEL"];
MakeSPheno[];
\end{lstlisting}
\item Leave \Mathematica, enter the \SPheno directory and create a new sub-directory for your model
\begin{lstlisting}
> cd $PATH/SPheno-3.3.0
> mkdir $MODEL
> cp $PATH/SARAH-4.5.0/Output/$MODEL/EWSB/SPheno/* $MODEL
\end{lstlisting}
\item Compile \SPheno together with the new module 
\begin{lstlisting}
> make Model=$MODEL 
\end{lstlisting}
\end{enumerate}
After these steps a new binary {\tt bin/SPheno\$MODEL} is available. To run it an input file in the Les Houches format is needed. 
\SARAH writes a template for that file which has to be filled with numbers. To enable the new functions for a calculation of the two-loop Higgs masses based on our new loop functions the following flags have to be set:
\begin{lstlisting}
Block SPhenoInput #
...
  7  0   # Skip two loop masses: True/False
  8  3   # Choose two-loop method
  9  1   # Gaugeless limit: True/False
\end{lstlisting}
{\tt 8 -> 3} chooses the new approach to calculate the loop corrections. The other options for flag {\tt 8} would correspond the effective potential calculations based on \SARAH ({\tt 8->1} for a fully numerical derivation, {\tt 8->2} for a semi-analytical derivation). Also some hard-coded corrections are available which are based on results in literature: {\tt 8->8} uses the known $\alpha_S(\alpha_b + \alpha_t)$ corrections for the MSSM, NMSSM, TMSSM or any variant thereof with up to four neutral CP-even Higgs fields and including models with Dirac gauginos \cite{Goodsell:DG}. {\tt 8->9} uses the corrections of option 8 and adds the two-loop MSSM $(\alpha_t + \alpha_b + \alpha_\tau)^2$ results based on Refs.~\cite{Brignole:2001jy,Degrassi:2001yf,Brignole:2002bz,Dedes:2002dy,Dedes:2003km}. Note that the last two options are not included by default in the \SPheno output of \SARAH. To include them, the user must make sure to include in the {\tt SPheno.m} of the considered model
\begin{lstlisting}
Use2LoopFromLiterature = True;
\end{lstlisting}

Finally, \SPheno is executed by
\begin{lstlisting}
> ./bin/SPheno$MODEL $MODEL/LesHouches.in.$MODEL 
\end{lstlisting}
and the output is written to
\begin{lstlisting}
 SPheno.spc.$MODEL
\end{lstlisting}
%%%$%%%%%% 

\subsection{Validation}

We have intensively used the \SPheno output to validate our new two-loop functions, in particular:
\begin{itemize}
 \item We found a numerical agreement of more than $10$ digits between our code and using public routines for the MSSM based on Refs.~\cite{Brignole:2001jy,Degrassi:2001yf,Brignole:2002bz,Dedes:2002dy,Dedes:2003km} for the self-energies. In order to perform this validation, it is necessary to use the same assumptions: turn off the first and second generation Yukawa couplings; take the Goldstone boson and light Higgs masses in the loops to be zero, and set the tree-level mixing angle of the neutral CP-even scalars to $\alpha=\beta-\pi/2 $. \\
The excellent agreement between all four possibilites to calculate the two-loop Higgs masses in the CMSSM is also shown in Fig.~\ref{fig:CMSSM}.The parameter point used here is the same as in \cite{Goodsell:2014bna},
  \begin{equation}
    \label{eq:CMSSMpoint}
    M_0 = M_{1/2} = 1~\TeV,\quad A_0=-2~\TeV,\quad \tan\beta=10,\quad \text{sgn}(\mu)=1
  \end{equation}

 \begin{figure}[hbt]
  \includegraphics[width=0.75\linewidth]{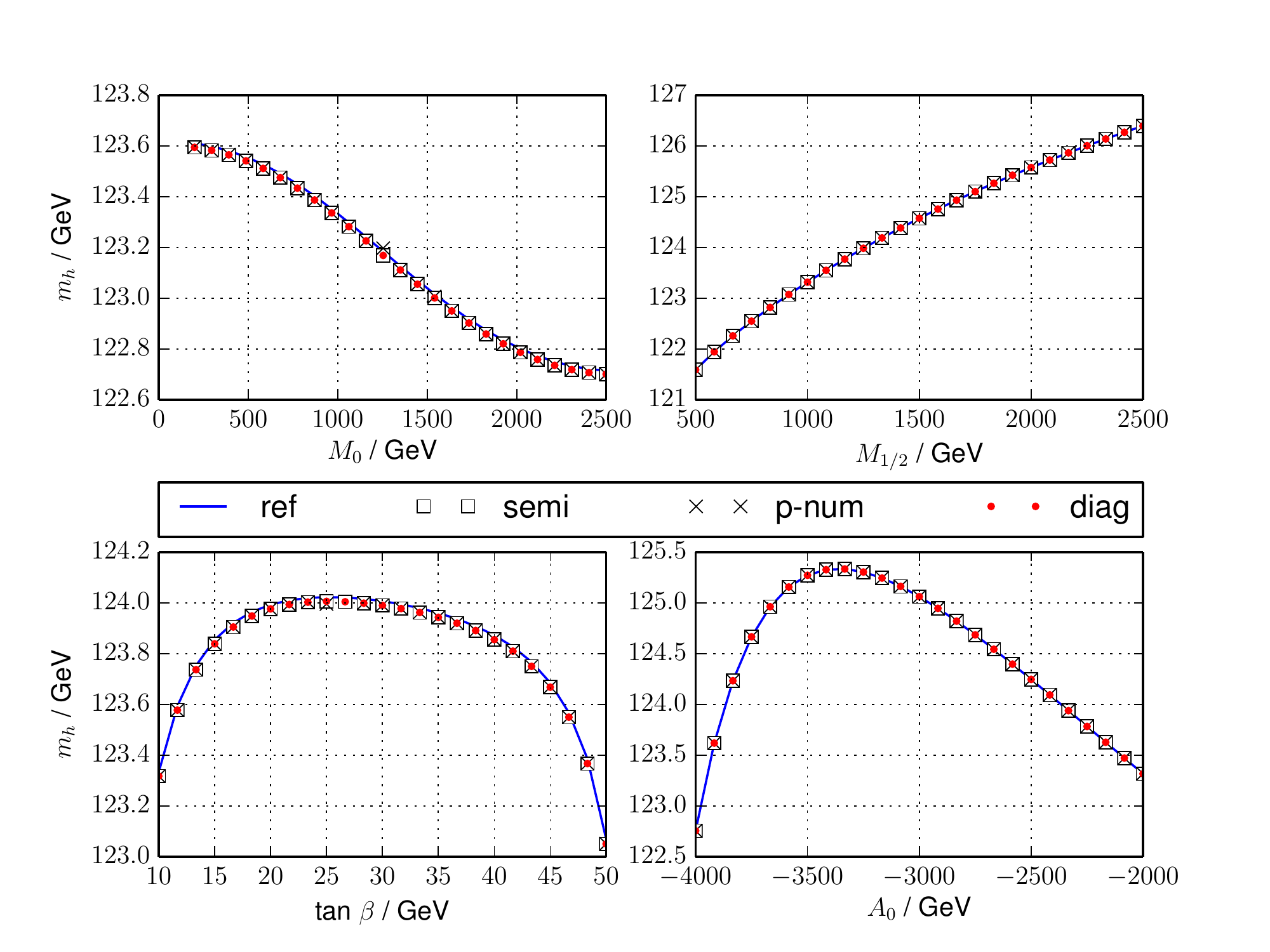}
 \caption{Comparison between the diagrammatic calculation of the two-loop Higgs masses with \SARAH/\SPheno presented here ({\tt diag}: diagrammatical) with the effective potential calculation of Ref.~\cite{Goodsell:2014bna} ({\tt p-num}: purely numerical, {\tt semi}: semi-analytical): , and with the routines based on Refs.~\cite{Brignole:2001jy,Degrassi:2001yf,Brignole:2002bz,Dedes:2002dy,Dedes:2003km} ({\tt ref}: reference). The fixed parameters are those in \qref{eq:CMSSMpoint}.}
 \label{fig:CMSSM}
 \end{figure}
 \item Similarly, we found full agreement with available results for the $\alpha_s(\alpha_b+\alpha_t)$ corrections in the NMSSM \cite{Degrassi:2009yq} and for Dirac gauginos \cite{Goodsell:DG}. 
 \item We compared the full two-loop corrections, i.e. also including corrections not involving the strong interaction, for the NMSSM, models with Dirac gauginos as well as for the B-L-SSM \cite{O'Leary:2011yq} against the results using the other two options based on a completely independent implementation in \SARAH presented in Ref.~\cite{Goodsell:2014bna}. We usually found very good agreement. Tiny differences were based on numerical artefacts in the routines using the effective potential ansatz. Similarly, we could reproduce the results of Ref.~\cite{Dreiner:2014lqa} for the two-loop contributions to the Higgs mass stemming from $R$-parity violating couplings. 
\end{itemize}
The new routines of course provide better stability. For example, in the routines based on numerically taking the derivatives of the potential, it is necessary to take care with the initial step size; if there are neutral scalars which have small expectation values then the results from those methods could become inaccurate -- this problem occurs in general for any neutral scalar having expectation value  $v_i \ll M_{SUSY}$. Less significantly, the numerical method can suffer (small) errors when there are small couplings present, such that they do not induce a sufficient shift in particle masses or couplings upon variation of the Higgs vevs to accurately take the derivative. Hence, it is very important to have two independent implementations of generic two-loop Higgs mass calculations in \SARAH/\SPheno: this is the only possibility to cross check results for models beyond the (N)MSSM at the moment. Thus, we highly motivate users to test all options for the model under consideration and to compare the results.

\section{Summary}
\label{sec:summary}
We have presented the derivation of a new set of expressions for calculation of the tadpoles and self-energies at the two-loop level. These expressions include all generic diagrams which do not vanish in the gaugeless limit and are valid in the limit of zero external momenta. This set of loop functions is simpler than the set of expressions obtained by taking the limit $p^2 \to 0$  in the pole mass functions available in literature so far. This allows for a rapid numerical evaluation of the Higgs mass. We have implemented these functions in \Fortran and included them in the new version of \SARAH\ {\tt 4.5.0}. This provides the possibility to automatically calculate the Higgs mass in a wide range of supersymmetric models with a guaranteed numerical accuracy and stability. The obtained precision for the Higgs mass is comparable with the one dedicated spectrum generators provide so far for the MSSM, and can now be applied to the study of a wide variety of models. 

Aside from accuracy and stability, one of the principal advantages of this approach is that it is readily extendable. It would be straightforward to extend the calculation to non-zero momentum by changing the functions in the code and linking with the library {\tt TSIL}. On the other hand, including the electroweak contributions should be possible by applying these techniques to the full effective potential; we presented the expressions in this case for the tadpoles in the appendix, but the second derivatives are currently unknown -- as are the full set of equivalent expressions in the diagrammatic approach. Furthermore, to truly reach the full two-loop precision we would require the two-loop shift in the Z-mass that determines the electroweak expectation value. We hope to return to these issues in future.

\section*{Acknowledgement}
We are indebted to Pietro Slavich for a large number of fruitful discussions, and comments on the draft.

\appendix

\section{Loop Functions}
\label{APP:loopfunctions}
The pole masses are constructed from various one- and two-loop functions which are defined in Ref.~\cite{Martin:2003qz}. However, for the purposes of calculating the effective potential these can be reduced to combinations of standard expressions involving just the usual functions $I$ and $J$. Here we compile this dictionary. We stick closely to the notation of Ref.~\cite{Martin:2003qz} and make often use of results presented there. Note that we use the standard notation
\begin{align}
\lnbar\ x \equiv \log \frac{x}{Q^2}
\end{align}
where $Q$ is the renormalisation scale.

\subsubsection{One loop functions}

At the one-loop level only two functions are needed:
\begin{align}
A(x) \equiv& \underset{{\epsilon \rightarrow 0}}{\mathrm{lim}} [ \mathbf{A}(x) + x/\epsilon ] = x( \ov{\ln} x -1) \nn\\
B(x,y) \equiv& \underset{{\epsilon \rightarrow 0}}{\mathrm{lim}} [ \mathbf{B}(x,y)-1/\epsilon]  = - \int_0^1 dt \ov{\ln}[tx + (1-t)y  - t(1-t) s]
\end{align}
Clearly, we find the following relation to $J(x)$ which is widely used at two loops: (recall $C\equiv 16\pi^2 \mu^{2\epsilon}(2\pi)^{-d}$)
\begin{align}
A(x) =& J(x),\quad {{\bf A}(x)={\bf J}(x)=J(x)-\frac x{\epsilon}} \label{eq:oneloopfunctionsAJ} \\
 \mathbf{B}(x,y) \underset{p^2 \rightarrow 0}\longrightarrow&  C \int d^dk \frac{1}{(k^2 + x)(k^2 + y)} = \frac{1}{y-x} ( {\bf J}(x) - {\bf J} (y)) \nn\\
B(x,y) \underset{p^2 \rightarrow 0}\longrightarrow& \frac{1}{y-x} ( J(x) - J (y)) \equiv B_0 (x,y)
\end{align}
where we have introduced the subscript to denote that the external momentum is zero. Let us also denote for future use
\begin{align}
C_0 (x,y,z) \equiv& \frac{B_0(x,z) - B_0(x,y)}{y-z}
\end{align}
which is actually symmetric on all three indices.

\subsubsection{Two-loop functions}
At the two-loop level we make use of the following set of functions
\begin{align}
J(x,y) \equiv& J(x) J(y) \\
S_0(x,y,z) =& I(x,y,z) \\
T_0(x,y,z) =& - \frac{\partial}{\partial x} I(x,y,z) \\
U_0(x,y,z,u) =& \frac{1}{y-x} ( I(x,z,u) - I (y,z,u)) \\
V_0(x,y,z,u) =& - \frac{\partial}{\partial y}U_0(x,y,z,u) \\
\ov{T}_0 (x,y)=& \underset{{\delta \rightarrow 0}}{\mathrm{lim}} [ T_0 (\delta,x,y) + B_0 (x,y) \ov{\ln} \delta]
\end{align}
together with
\begin{align}
 \mathbf{M}(x,y,z,u,v) \rightarrow &C^2 \int d^d k \int d^dq \frac{1}{(k^2+x)(q^2+y)(k^2 +z)(q^2+u)((k-q)^2+v)} \nn\\
=& \frac{1}{(u-y)} (\mathbf{U} (x,z,y,v) - \mathbf{U} (x,z,u,v)) \nn\\
M_0(x,y,z,u,v) =& \frac{1}{(u-y)} (U_0 (x,z,y,v) - U_0 (x,z,u,v)).
\end{align}
These functions have to following properties: (i) $I(x,y,z)$ is symmetric in all arguments; (ii) $T_0(x,y,z)$ is symmetric in the last two arguments; (iii) $U_0(x,y,z,u)$ is invariant under the exchange $z \leftrightarrow u$ and $x \leftarrow y$; (iv) $M_0(x,y,z,u,v)$ is invariant under the interchanges 
$(x,z) \leftrightarrow (y,u)$, $(x,y) \leftrightarrow (z, u)$, and $(x,y) \leftrightarrow (u, z)$.

An explicit expression for $I(x,y,z)$ is for instance given by \cite{Martin:2001vx}
\begin{align}
I(x,y,z) =& \frac{1}{2} (x-y-z) \ov{\ln}y \ov{\ln}z + \frac{1}{2}(y-x-z)\ov{\ln}x\ov{\ln}z + \frac{1}{2}(z-x-y)\ov{\ln}x\ov{\ln}y  \nn \\
 & + 2 x \ov{\ln}x + 2 y \ov{\ln}y + 2z \ov{\ln}z - \frac{5}{2} (x+y+z) - \frac{1}{2} \xi(x,y,z)
\end{align}
with
\begin{align}
 \xi(x,y,z) =& R \Big[ 2 \ln[(z+x-y-R)/2z] \ln[(z+y-x-R)/2z] - \ln\frac{x}{z} \ln\frac{y}{z}  \nn \\
 & - 2 \text{Li}_2[(z+x-y-R)/2z] - 2 \text{Li}_2[(z+y-x-R)/2z] + \frac{\pi^2}{3} \Big] \\
 R =& \sqrt{x^2 + y^2 + z^2 - 2xy -2xz -2yz} .
\end{align}
Note that $x,y \leq z$ has been assumed here.

\subsubsection{Relations required for the pole functions}

The above loop functions are used as a basis for the various loop functions. However, we also find additional combinations such as
\begin{align}
V_{SSSSS} (x,y,z,u,v) =& \frac{1}{y-z} (U_0 (x,y,u,v) - U_0 (x,z,u,v)) \nn\\
\sim& - \frac{1}{(k^2 +x)(k^2 + y)(k^2+z)} \frac{1}{(q^2 + u)((q+k)^2+v)}.
\end{align}
we should compare this to
\begin{align}
V(x,y,z,u) \sim& \frac{1}{(k^2 +x)(k^2 + y)^2} \frac{1}{(q^2 + u)((q+k)^2+v)}.
\end{align}
Hence we can write $V_{SSSSS} (x,y,y,u,v) = - V(x,y,z,u)$ and remember that $V_{SSSSS}$ is symmetric on its first three entries.

\subsubsection{Simplified loop functions}

For the amplitudes with one or more massless (and possibly identical) fields we find simplified expressions for the loop integrals, some of which are collected below.
\begin{align}
I(x,y,0) =& \frac{1}{2}(-5x - 5y + (-x + y)\lnbar^2 x + 4y\lnbar y +   \lnbar x (4x - 2y\lnbar y)-2(x - y)\mathrm{Li}_2 (1 - y/x)) \\
I(x,x,0) =& x(-5 + 4 \lnbar x - \lnbar^2 x) \\
I(x,0,0) =& - \frac{1}{2} x (\ov{\ln}x)^2 + 2x\ov{\ln}x - \frac{5}{2} x - \frac{\pi^2}{6}x \\
B_0 (x,x) =& - \lnbar x 
\end{align}
It is also sometimes necessary to consider the case of small mass splittings. The results for $I(\delta,x,y)$, $I(\delta,x,x)$, $I(\delta,\delta',x)$ in the limits $\delta \to 0$, $\delta' \to 0$ can be found in Ref.~\cite{Martin:2001vx} and we do not repeat them here.

\section{Second derivatives of the effective potential}
\label{APP:SecondDerivatives}

In this appendix we present the results for the second derivatives of the effective potential. Largely these are identical to those in \cite{Martin:2003it} with the external momentum set to zero, but for sake of completeness we repeat the full set here. However, certain expressions become much simpler in this limit, notably some complicated functions involving both fermion and scalar propagators, and those involving gauge bosons. 

The full contribution is
\begin{align}
\Pi_{ij} =&  \Pi^{S}_{ij} + \Pi^{SF(W)}_{ij} + \Pi^{SF_4 (M)}_{ij} + \Pi^{S_2 F_3 (M)}_{ij} + \Pi^{S_3F_2 (V)}_{ij} + \Pi^{SF_4 (V)}_{ij} + \Pi^{SV}_{ij} + \Pi_{ij}^{FV}.
\end{align}

\subsection{Diagrams with only scalar propagators}

The first contribution, including only scalar propagators, comprises the eight diagrams shown in \ref{fig:SelfScalars}. These are unchanged from \cite{Martin:2003it} and are given by
\begin{eqnarray} 
&& \Pi^{S}_{ij} \>=\>
\frac{1}{4} \lambda^{ijkl} \lambda^{kmn} \lambda^{lmn}
\propW_{SSSS} (m_k^2, m_l^2, m_m^2, m_n^2)
+ \frac{1}{4} \lambda^{ijkl} \lambda^{klmm} \propX_{SSS}
(m_k^2, m_l^2, m_m^2)
\nonumber \\ && \quad
+ \frac{1}{2} \lambda^{ikl} \lambda^{jkm} \lambda^{lmnn} 
\propY_{SSSS} (m_k^2, m_l^2, m_m^2, m_n^2)
+ \frac{1}{4} \lambda^{ikl} \lambda^{jmn} \lambda^{klmn}
\propZ_{SSSS} (m_k^2, m_l^2, m_m^2, m_n^2)
\nonumber \\ && \quad
+ \frac{1}{6} \lambda^{iklm} \lambda^{jklm} \propS_{SSS} 
(m_k^2,m_l^2,m_m^2) 
+\frac{1}{2} \left (
\lambda^{ikl} \lambda^{jkmn} + \lambda^{jkl} \lambda^{ikmn} \right )
\lambda^{lmn} \propU_{SSSS}
(m_k^2, m_l^2, m_m^2, m_n^2) 
\nonumber \\ && \quad
+ \frac{1}{2} \lambda^{ikl} \lambda^{jkm} \lambda^{lnp} \lambda^{mnp} 
\propV_{SSSSS} (m_k^2, m_l^2, m_m^2, m_n^2, m_p^2)
+ \frac{1}{2} \lambda^{ikm} \lambda^{jln} \lambda^{klp} \lambda^{mnp}
\propM_{SSSSS}(m_k^2, m_l^2, m_m^2, m_n^2, m_p^2) .
\end{eqnarray}
Here the loop integral functions 
are given by:
\begin{eqnarray}
\propW_{SSSS}(x,y,z,u) &=& U_0(x,y,z,u),
\\
\propX_{SSS}(x,y,z) &=& - J(z) B_0(x,y) ,
\\
\propY_{SSSS}(x,y,z,u) &=& J(u) C_0(x,y,z) ,
\\
\propZ_{SSSS}(x,y,z,u) &=& B_0(x,y) B_0(z,u) ,
\\
\propS_{SSS}(x,y,z) &=& -I(x,y,z) ,
\\
\propU_{SSSS}(x,y,z,u) &=& U_0(x,y,z,u) ,
\\
\propV_{SSSSS}(x,y,z,u,v) &=& [U_0(x, y, u, v) -U_0(x, z, u, v)]/(y - z) ,
\\
\propM_{SSSSS}(x,y,z,u,v) &=& -M_0(x,y,z,u,v) .
\end{eqnarray}
In the case that $y=z$, we have the simplification
\begin{align}
V_{SSSSS}(x,y,y,u,v) =& - V(x,y,u,v).
\end{align}

It should be noted (for example, for the purposes of evaluating the colour factors) that topologies $X_{SSS}, Y_{SSS}, Z_{SSS}$ arise from differentiating $V_{SS}^{(2)}$ (and hence the tadpole $T_{SS}$) while the others arise from differentiating $V_{SSS}^{(2)}$ (and hence the tadpoles $T_{SSS}$ and $T_{SSSS}$).

\subsection{Diagrams with scalar and fermion propagators}

The contributions from diagrams with the topology $\propW$ are
\begin{eqnarray}
\Pi^{SF(W)}_{ij} &=&
\frac{1}{2} \lambda^{ijkl}
\re \bigl [y^{MNk} y^{M'N'l} M_{MM'} M_{NN'} \bigr ]
\propW_{SS\Fbar\Fbar} (m_k^2, m_l^2, m_M^2, m_N^2)
\nonumber \\ &&
+\frac{1}{2} \lambda^{ijkl} y^{MNk} y_{MNl}
\propW_{SSFF} (m_k^2, m_l^2, m_M^2, m_N^2),
\end{eqnarray}
where we can slightly simplify the loop functions:
\begin{eqnarray}
\propW_{SS\Fbar\Fbar}(x,y,z,u) &=& -2 \propW_{SSSS}(x,y,z,u)
,
\\
\propW_{SSFF}(x,y,z,u) &=& -(z+u-y) U_0(x,y,z,u)-I(x,z,u)+B_0(x,y) (J(z)+J(u)).
\end{eqnarray}
The contributions from diagrams of the topology $\propM$ with four fermions are
\begin{eqnarray}
\Pi^{SF_4 (M)}_{ij} &=&
\re \bigl [ y^{KMi} y^{LNj} y^{K'L'p} y^{M'N'p}
M_{KK'} M_{LL'} M_{MM'} M_{NN'} \bigr ]
\propM_{\Fbar\Fbar\Fbar\Fbar S}(m_K^2, m_L^2, m_M^2, m_N^2, m_p^2)
\nonumber \\ &&
+ 2 \re \bigl [ y^{KMi} y_{LNj} y_{KL'p} y^{M'Np} M^{LL'} M_{MM'} \bigr ]
\propM_{F\Fbar\Fbar FS}(m_K^2, m_L^2, m_M^2, m_N^2, m_p^2)
\nonumber \\ && 
+ \re \bigl [
\left ( y^{KMi} y_{LNj} + y^{KMj} y_{LNi} \right )
y_{KL'p} y_{MN'p} M^{LL'} M^{NN'} \bigr ]
\propM_{F\Fbar F\Fbar S}(m_K^2, m_L^2, m_M^2, m_N^2, m_p^2)
\nonumber \\ && 
+ 2 \re \bigl [ y^{KMi} y^{LNj} y_{KLp} y^{M'N'p} M_{MM'} M_{NN'} \bigr ]
\propM_{FF\Fbar\Fbar S}(m_K^2, m_L^2, m_M^2, m_N^2, m_p^2)
\nonumber \\ &&
+ \re \bigl [ y^{KMi} y^{LNj} y_{KLp} y_{MNp} \bigr ]
\propM_{FFFFS}(m_K^2, m_L^2, m_M^2, m_N^2, m_p^2) ,
\end{eqnarray}
where
\begin{eqnarray}
\propM_{\Fbar\Fbar\Fbar\Fbar S}(x,y,z,u,v) &=& 2 M_0(x,y,z,u,v) ,
\\
\propM_{F\Fbar \Fbar FS}(x,y,z,u,v) &=& 
(y+z-v) M_0(x,y,z,u,v) 
- U_0(x,z,u,v) 
- U_0(u,y,x,v) 
+ B_0(x,z) B_0(y,u) ,
\phantom{feif}
\\
\propM_{F\Fbar F\Fbar S}(x,y,z,u,v) &=& 
(x+z) M_0(x,y,z,u,v) - U_0(y,u,z,v) - U_0(u,y,x,v) ,
\\
\propM_{FF\Fbar\Fbar S}(x,y,z,u,v) &=& (x+y-v) M_0(x,y,z,u,v)
- U_0(x, z, u, v) - U_0(y, u, z, v) 
+ B_0(x,z) B_0(y, u)  ,
\\
\propM_{FFFFS}(x,y,z,u,v) &=&
(x u +y z) M_0(x,y,z,u,v) 
-x U_0(z,x,y,v) -z U_0(x,z,u,v)
\nonumber \\ && 
-u U_0(y,u,z,v)
-y U_0(u,y,x,v)
+ I(x,u,v) + I(y,z,v).
\end{eqnarray}
The results from diagrams of the topology $\propM$ with three fermions are
\begin{eqnarray}
\Pi^{S_2 F_3 (M)}_{ij} &=&
\lambda^{ikm} 
\Bigl (
\re \bigl [ y^{LNj} y^{L'Pk} y^{N'P'm} M_{LL'} M_{NN'} M_{PP'} \bigr ]
\propM_{S \Fbar S \Fbar \Fbar}(m_k^2, m_L^2, m_m^2, m_N^2, m_P^2)
\nonumber \\ && 
+
2 \re \bigl [ y^{LNj} y_{LPk} y^{N'Pm} M_{NN'} \bigr ]
\propM_{S F S \Fbar F}(m_k^2, m_L^2, m_m^2, m_N^2, m_P^2)
\nonumber \\ && 
+
\re \bigl [y^{LNj} y_{LPk} y_{NP'm} M^{PP'} \bigr ]
\propM_{S F S F \Fbar}(m_k^2, m_L^2, m_m^2, m_N^2, m_P^2)
\Bigr ) 
+ (i \leftrightarrow j) ,
\end{eqnarray}
where
\begin{eqnarray}
\propM_{S \Fbar S \Fbar \Fbar}(x,y,z,u,v) &=& 2 M_0(x,y,z,u,v) ,
\\
\propM_{S F S \Fbar F}(x,y,z,u,v) &=& (v - x +y) M_0(x,y,z,u,v)
+ U_0(y,u,z,v) - U_0(x,z,u,v) 
- B_0(x,z) B_0(y,u) 
,
\\
\propM_{S F S F \Fbar}(x,y,z,u,v) &=& 
(y+u) M_0(x,y,z,u,v) - U_0(x, z, u, v) - U_0(z, x, y, v) 
.
\end{eqnarray}

The contributions from diagrams of topology $\propV$, with three scalars and two fermions are
\begin{eqnarray} 
\Pi^{S_3F_2 (V)}_{ij} &=&
\lambda^{ikl} \lambda^{jkm}
\Bigl ( 
\re \bigl [ y^{NPl} y^{N'P'm} M_{NN'} M_{PP'} \bigr ]
\propV_{SSS\Fbar\Fbar} (m_k^2, m_l^2, m_m^2, m_N^2, m_P^2)
\nonumber \\ && 
+ 
\re \bigl [y^{NPl} y_{NPm} \bigr ]
\propV_{SSSFF} (m_k^2, m_l^2, m_m^2, m_N^2, m_P^2)
\Bigr )
,
\end{eqnarray}
where
\begin{eqnarray}
\propV_{SSS\Fbar\Fbar}(x,y,z,u,v) &=&
-2 \propV_{SSSSS}(x,y,z,u,v) 
,
\\
\propV_{SSSFF}(x,y,z,u,v)
&=& U_0(x,y,u,v) +(z-u-v)V_{SSSSS}(x,y,z,u,v)-(J(u) +J(v)) C_0(x,y,z).
\end{eqnarray}

The results from diagrams of topology $\propV$ with four fermions are
\begin{eqnarray}
\Pi^{SF_4 (V)}_{ij} &=&
2 \re \bigl [
y^{KLi} y^{K'Mj} y^{L'Np} y^{M'N'p} M_{KK'} M_{LL'} M_{MM'} M_{NN'} 
\bigr ]
\propV_{\Fbar\Fbar\Fbar\Fbar S} (m_K^2, m_L^2, m_M^2, m_N^2, m_p^2) 
\nonumber \\ && 
+ 2 \re \bigl [ 
\bigl (y^{KLi} y^{K'Mj} +y^{KLj} y^{K'Mi} \bigr ) 
y_{LNp} y^{M'Np} M_{KK'} M_{MM'} 
\bigr ]
\propV_{\Fbar F\Fbar FS} (m_K^2, m_L^2, m_M^2, m_N^2, m_p^2)
\nonumber \\ &&
+ 2 \re \bigl [ y^{KLi} y^{K'Mj} y_{LNp} y_{MN'p} M_{KK'} M^{NN'} \bigr ]
\propV_{\Fbar FF \Fbar S} (m_K^2, m_L^2, m_M^2, m_N^2, m_p^2) 
\nonumber \\ &&
+ 2 \re \bigl [ y^{KLi} y_{KMj} y^{L'Np} y_{M'Np} M_{LL'} M^{MM'} \bigr ]
\propV_{F\Fbar\Fbar F S} (m_K^2, m_L^2, m_M^2, m_N^2, m_p^2)
\nonumber \\ &&
+ 2 \re \bigl [
\left ( y^{KLi} y_{KMj} +y^{KLj} y_{KMi} \right )
y_{LNp} y_{M'N'p} M^{MM'} M^{NN'} \bigr ]
\propV_{FF\Fbar\Fbar S} (m_K^2, m_L^2, m_M^2, m_N^2, m_p^2)
\nonumber \\ &&
+ 2 \re \bigl [
y^{KLi} y_{KMj} y_{LNp} y^{MNp} \bigr ]
\propV_{FFFFS} (m_K^2, m_L^2, m_M^2, m_N^2, m_p^2)
,
\end{eqnarray}
where
\begin{align}
\propV_{\Fbar\Fbar\Fbar\Fbar S}(x,y,z,u,v) =& -2 \propV_{SSSSS}(x,y,z,u,v)
,
\\
\propV_{\Fbar F\Fbar FS}(x,y,z,u,v) =& -U_0(x,y,u,v) +(v-z-u)V_{SSSSS}(x,y,z,u,v)-(J(v) -J(u)) C_0(x,y,z)
\\
\propV_{\Fbar FF\Fbar S}(x,y,z,u,v) =& -2 U_0(x,y,u,v) -2z V_{SSSSS} (x,y,z,u,v)
\\
\propV_{F\Fbar\Fbar FS}(x,y,z,u,v)=& f^{(2,0,0)}_{FFS} (x,y,z; u,v) 
\\
\propV_{FF\Fbar\Fbar S}(x,y,z,u,v) =&  -U_0(x,y,u,v) - U_0(y,z,u,v) -(x+z) V_{SSSSS}(x,y,z,u,v)
\\
\propV_{FFFFS}(x,y,z,u,v) =& f^{(1,0,0)}_{FFS} (y,z, u; v) +x f^{(2,0,0)}_{FFS} (x,y,z, u;v) .
\end{align}
These represent significant simplifications over the full pole contributions. To recapitulate,
\begin{align}
f^{(1,0,0)}_{FFS} (x,y,u; v) \equiv& B_0(x,y) (J(v)-J(u))+I(x,u,v)-(y+u-v)U_0(x,y,u,v)\nn\\
f^{(2,0,0)}_{FFS} (x,y,z,u;v) \equiv& C_0(x,y,z) (J(u)-J(v))-U_0(x,z,u,v)-(y+u-v) V_{SSSSS}(x,y,z,u,v).
\end{align}
Note that $ f^{(2,0,0)}_{FFS}$ is symmetric on its first three indices. 

\subsection{Diagrams with one vector propagator}
For self energies of neutral scalars where all gauge groups are unbroken, the diagrams involving one vector propagator are particularly simple.

\subsubsection{Diagrams with scalars}

 We have for diagrams involving scalars
\begin{align}
\Pi^{SV}_{ij} =& \frac{1}{2} g^2 d(i) C(i) \bigg[ \lambda^{ijkk} W_{SSSV} (m_k^2, m_k^2, m_k^2,0) + \lambda^{ikl} \lambda^{jkl} G_{SS} (m_k^2, m_l^2) \bigg].
\end{align}
In \cite{Martin:2003it} the functions are given as (setting the external momentum to zero) 
\begin{align}
W_{SSSV} ( x,x,x,0) \equiv& 3 I(x,x,0) - J(x) + 2x \\
G_{SS} (x,y) \equiv& 4 y V(x, y,y,0) + 4 x V(y, x,x,0) - 2U_0(x,y,y,0) - 2U_0(y,x,x,0) - 2J(y) B_0(x, y') - 2J(x) B_0(y,x') \nn\\
&+ 2(x+y) M (x,x,y,y,0) -   2U_0(x,y,y,0) - 2U_0(y,x,x,0) + B_0(x,y)^2.
\end{align}
However, the expression for $G_{SS}$ greatly simplifies, as we could see by taking the derivative of (\ref{EQ:TadpoleScalarVector}):
\begin{align}
G_{SS} (x,y) =& 2 \bigg[  -U_0(m_i^2,m_k^2,m_k^2,0)- U_0(m_m^2,m_i^2,m_i^2,0) + B_0 (m_i^2, m_k^2) + 2) \bigg] \nn\\
=& - 12 + \frac{11 (x \lnbar x - y \lnbar y) - 3 (x \lnbar^2 x - y \lnbar^2 y)}{x-y}\ , \\
G_{SS} (x,x) =& -1 + 5 \lnbar x - 3 \lnbar^2 x \ .
\end{align}

\subsubsection{Diagrams with fermions}

For the diagrams involving fermions we obtain
\begin{align}
\Pi_{ij}^{FV} =& g^2 d(K) C(K) \big[\mathrm{Re}(y^{iKL} y_{jKL}) G_{FF} (m_K^2,m_L^2) + \mathrm{Re}(y^{ i KL} y^{j K'L'}M_{KK'} M_{LL'}  ) G_{\ov{FF}} (m_K^2,m_L^2)\big].
\end{align}
Here we have the simpler expressions
\begin{align}
G_{FF} (x,y) \equiv& 2 (x+y) [ 3U_0(x,y,x,0) + 3 U_0(x,y,y,0) - 5B_0(x,y)] - 6I(x,x,0) - 6I(y,y,0) + 10 J(x) + 10 J(y) - 16 (x+y)  \nn\\
& + \delta_{\ov{\mathrm{MS}}} 4\big[ J(x) + J(y) - (x+y)B_0(x,y) \big] \nn\\
 G_{\ov{FF}} (x,y) \equiv& 4 \bigg( 3 U_0(x,y,x,0) + 3 U_0(x,y,y,0) - 5 B_0(x,y) -4 \bigg) \nn\\
& - \delta_{\ov{\mathrm{MS}}} 4\big[ 1 + 2 B_0(x,y)\big]
\end{align}

\section{Derivatives of effective potential with massive vectors}
\label{APP:full}

In this appendix we shall present the full results for the tadpoles including the possibility of massive gauge bosons. To this end, instead of parametrising the gauge interactions via covariant derivatives, we shall instead use the notation of \cite{Martin:2001vx} and supplement our interactions (\ref{EQ:couplingdefinitions}) with
\begin{align}
{\cal L}_{\rm SV} =& 
- g^{aij} A^\mu_a \phi_i \partial_\mu \phi_j
-\frac{1}{4} g^{abij}  A^\mu_a A_{\mu b} \phi_i \phi_j
-\frac{1}{2} g^{abi} A^\mu_a A_{\mu b} \phi_i
,
\\
{\cal L}_{\rm FV} =& g^{aJ}_I A_a^\mu \psi^{\dagger I} {\overline
\sigma}_\mu \psi_J
\end{align}
with repeated indices summed over, and metric signature ($-$$+$$+$$+$). 

\subsection{Effective potential}

The full two-loop effective potential in the Landau gauge was given in Ref.~\cite{Martin:2001vx}:
\begin{align}
V^{(2)}_{SSS} =& \frac{1}{12} (\lambda^{ijk})^2 f_{SSS} (m_i^2, m_j^2,m_k^2) \\
V^{(2)}_{SS} =& \frac{1}{8} \lambda^{iijj} f_{SS} (m_i^2, m_j^2) \\
V^{(2)}_{FFS} =& {1\over 2} |y^{IJk}|^2 f_{FFS} (m^2_I, m^2_J, m^2_k) ,\\  
V^{(2)}_{\ov{FF}S} =& {1\over 4} y^{IJk} y^{I'J'k} M^*_{II'} M^*_{JJ'} 
                             f_{\ov{FF}S} (m^2_I, m^2_J, m^2_k) + {\rm c.c.},\\
V^{(2)}_{SSV} =& {1\over 4} (g^{aij})^2 f_{SSV} (m^2_i,m^2_j, m_a^2),\\
V^{(2)}_{VS} =& {1\over 4} g^{aaii} F_{VS} (m^2_a, m^2_i) 
 ,\\ 
V^{(2)}_{VVS} =& {1\over 4} (g^{abi})^2 f_{VVS} (m_a^2, m_b^2,m^2_i) ,\\
V^{(2)}_{FFV} =& {1\over 2} |g^{aJ}_I|^2 f_{FFV} (m^2_I, m^2_J, m_a^2) ,\\
V^{(2)}_{\ov{FF}V} =& {1\over 2} g^{aJ}_I g^{aJ'}_{I'} M^{II'} M^*_{JJ'} f_{\ov{FF}V} (m^2_I, m^2_J, m_a^2),\\
V^{(2)}_{\mathrm{gauge}} =& \frac{1}{12} (g^{abc})^2 f_{\rm gauge} (m_a^2, m_b^2,m_c^2).
\end{align}
The modified loop functions are 
\begin{equation*}
\begin{array}{|c|c|c|} \hline 
 &\ov{\mathrm{DR}}^\prime  &\Delta_{\ov{\mathrm{MS}}}  \\ \hline
f_{SSV} & \frac{1}{z} \Big[ -\Delta(x,y,z) I(x,y,z) + (x-y)^2 I(0,x,y)& \\ 
& + (y-x-z) J(x,z) + (x-y-z)J(y,z) + z J(x,y)\Big]& 0 \\
&+ 2 (x+y-z/3) J(z) & \\ \hline
f_{VS} & 3J(x,y) & 2xJ(y)\\ \hline
f_{VVS} &\frac{1}{4xy}\big[ (-\Delta(x,y,z) - 12xy)I(x,y,z) & \\
&+ (x-z)^2 I(0,x,z) + (y-x)^2 I(0,y,z) - z^2 I(0,0,z) &2J(z) - x - y - z \\
&+(z-x-y)J(x,y)+yJ(x,z) + xJ(y,z) \big]& \\ 
& +\frac{1}{2} J(x) + \frac{1}{2} J(y)& \\ \hline
f_{FFV} &  \frac{1}{z} \big[ (\Delta(x,y,z)-3z^2 + 3 xz + 3 yz)) I(x,y,z) - (x-y)^2 I(0,x,y) &\\ 
&  + (x-y-2z) J(x,z) + (y-x-2z)J(y,z) + 2z J(x,y) \big]&-2xJ(x) - 2yJ(y) + (x+y)^2 -z^2\\
&  + 2 (-x -y+z/3) J(z) & \\ \hline
f_{\ov{FF}V} & 6I(x,y,z) & 2(x+y+z) -4J(x) -4J(y)\\ \hline
f_{\rm gauge} & {1\over 4 x y z} \Bigl \lbrace (-x^4 - 8 x^3 y - 8 x^3 z +32 x^2 y z +18 y^2 z^2 ) I(x,y,z)  &  \\
& + (y-z)^2 (y^2 + 10 y z + z^2) I (0,y,z) + x^2 (2 y z - x^2) I (0,0,x) & \\
& + (x^2 - 9 y^2 - 9 z^2  +9 x y + 9 x z + 14 y z) x J(y,z)& x^2 + 12yz + 2x J(x) \\
& + (22 y + 22 z- 40x/3) xyz J(x)\Bigr \rbrace + (x \leftrightarrow y) + (x \leftrightarrow z)  &  + (x \leftrightarrow y) + (x \leftrightarrow z)\\\hline
\end{array}
\end{equation*}
where $\Delta(x,y,z) \equiv x^2 + y^2 + z^2 - 2xy - 2xz-2yz$ and it is understood that $f_\alpha^{\ov{\mathrm{MS}}} = f_\alpha^{\ov{\mathrm{DR}}'} + \Delta_{\ov{\mathrm{MS}}}$.

\subsection{Tadpoles}

Clearly our results for the derivatives involving no vectors given in the text are unchanged. However, for all others we will have to apply our procedure to the more complicated loop functions and also the derivatives of the scalar masses and couplings. In fact, of the new couplings only $g^{abi}$ has a non-trivial derivative, so we require
\begin{align}
m_{ab}^2 (S) \delta^{\mu \nu} =& - \frac{\partial^2 \mathcal{L}}{\partial A^a_\mu \partial A^a_\nu}\nn\\
m_{ab}^2 (S)&= m_{a}^2 \delta^{ab} + g^{abi} S_i + \frac{1}{2} g^{abij} S_i S_j \nn\\
g^{abi} (S) &= g^{abi} + g^{abij} S_j,
\end{align}
and therefore
\begin{align}
\frac{\partial}{\partial S_r} g^{abi} =& g^{abir} \nn\\
\frac{\partial}{\partial S_r} m_{ab}^2 (S) =& g^{abr}(S)
\end{align}

In the following we define
\begin{align}
f_\alpha^{(1,0,0)} (x,u; y,z) \equiv& \frac{f_\alpha (x,y,z) - f_\alpha(u,y,z)}{x-u} \nn\\
f_\alpha^{(0,0,1)} (x,y; z,u ) \equiv& \frac{f_\alpha (x,y,z) - f_\alpha(x,y,u)}{z-u}
\end{align}
i.e. we can give the loop functions used for the tadpoles in terms of those in the effective potential. In general these can be simplified, in particular to allow the smooth limit $u \rightarrow x$ to be taken, but we postpone that to future work where we shall also treat the second derivatives. Here we simply present the full set of tadpole diagrams, modifying (\ref{EQ:FullTadpoleGaugeless}):
\begin{align}
\frac{\partial V^{(2)}}{\partial S_p^0} = R_{rp}^0 [T_{S} + T_{SSFF}+ T_{FFFS}+T_{SSV}+T_{VS}+T_{VVS}+T_{FFV}+T_{\Fbar\Fbar V}+ T_{\rm gauge}]
\end{align}
where $T_{S} , T_{SSFF},T_{FFFS} $ are as given in the body of the paper and 
\begin{align}
T_{SSV} =&  \frac{1}{2} g^{aij} g^{akj} \lambda^{ikr} f_{SSV}^{(1,0,0)} (m_i^2,m_k^2; m_j^2, m_a^2) + \frac{1}{4}  g^{aij} g^{bij} g^{abr} f_{SSV}^{(0,0,1)} (m_i^2, m_j^2; m_a^2,m_b^2)\\
T_{SSV} =& {1\over 4} g^{abii} g^{abr}  f_{VS}^{(1,0)} (m^2_a, m^2_b;m_i^2) + {1\over 4} g^{aaik} \lambda^{ikr}  f_{VS}^{(0,1)} (m^2_a; m^2_i,m_k^2) \\
T_{VVS} =& \frac{1}{2} g^{abi} g^{cbi} g^{acr} f_{VVS}^{(1,0,0)}( m_a^2, m_c^2; m_b^2,m_i^2) + {1\over 4} g^{abi} g^{abj} \lambda^{ijr}f_{VVS}^{(0,0,1)} (m_a^2, m_b^2;m^2_i,m_j^2)\\
T_{FFV} =&2 g^{aJ}_{I}  \ov{g}_{bJ}^{K} \mathrm{Re} ( M_{KI'} y^{I'Ir}) f_{FFV}^{(1,0,0)} (m^2_I, m_K^2; m^2_J, m_a^2) + {1\over 2} g^{aJ}_I \ov{g}_{bJ}^I g^{abr} f_{FFV}^{(0,0,1)} (m^2_I, m^2_J; m_a^2 ,m_b^2) \\
T_{\Fbar \Fbar V} =& g^{aJ}_I g^{aJ'}_{I'} \mathrm{Re} (y^{II'r} M^*_{JJ'}) \big[ f_{\ov{FF}V} (m^2_I, m^2_J, m_a^2) + M_I^2 f_{\ov{FF}V}^{(1,0,0)} (m^2_I, m_{I'}^2 ; m^2_J, m_a^2) \big]\nn\\
&+  g^{aJ}_I g^{aJ'}_{I'} \mathrm{Re}( M^{IK'} M^{KI'} M^*_{JJ'} y_{KK'r}) f_{\ov{FF}V}^{(1,0,0)} (m^2_I, m_{I'}^2 ; m^2_J, m_a^2)\nn\\
&+ \frac{1}{2} g^{aJ}_I g^{bJ'}_{I'} g^{abr} M^{II'} M^*_{JJ'} f_{\ov{FF}V}^{(0,0,1)} (m^2_I, m^2_J; m_a^2,m_b^2) \\
T_{\rm gauge} =& \frac{1}{4} g^{abc} g^{dbc} g^{adr} f_{\rm gauge}^{(1,0,0)} (m_a^2, m_d^2; m_b^2,m_c^2).
\end{align}

\bibliography{diagrammaticpaper.bbl}

\end{document}